\documentclass[reprint,amsmath,amssymb,pra,]{revtex4-2}

\usepackage{graphicx}
\usepackage{dcolumn}
\usepackage{bm}
\usepackage{quantikz}
\usepackage{amsmath,amssymb} 
\usepackage{amsthm,epsfig,tikz}
\usepackage[justification=justified]{caption}
\usepackage{xcolor}
\usepackage{verbatim}
\usepackage[normalem]{ulem}
\usepackage[rightcaption]{sidecap}
\usepackage{subfigure}
\usepackage{forest,adjustbox}
\usepackage{multirow}
\usepackage{hyperref}  
\hypersetup{colorlinks,citecolor=black,urlcolor=black,linkcolor =black,hypertexnames=true,pdfproducer={Me}}
\usepackage{tikz}
\usepackage{url}

\usepackage{float}
\makeatletter
\let\newfloat\newfloat@ltx
\makeatother
\usepackage{algorithm}
\usepackage{algcompatible}

\newcommand{\subf}[2]{%
  {\small\begin{tabular}[t]{@{}c@{}}
  #1\\#2
  \end{tabular}}%
}

\begin{document}

\title{Empirical Power of Quantum Encoding Methods for Binary Classification}
\author{Gennaro De Luca}
 \affiliation{School of Computing and Augmented Intelligence \\ 
 Arizona State University
 }
 
\author{Andrew Vlasic}
\affiliation{
Deloitte Consulting, LLP
}

\author{Michael Vitz}
\affiliation{
Deloitte Consulting, LLP
}

\author{Anh Pham}
\affiliation{
Deloitte Consulting, LLP
}

\date{\today}

\begin{abstract}
Quantum machine learning is one of the many potential applications of quantum computing, each of which is hoped to provide some novel computational advantage. However, quantum machine learning applications often fail to outperform classical approaches on real-world classical data. The ability of these models to generalize well from few training data points is typically considered one of the few definitive advantages of this approach. In this work, we will instead focus on encoding schemes and their effects on various machine learning metrics. Specifically, we focus on real-world data encoding to demonstrate differences between quantum encoding strategies for several real-world datasets and the classification model standard, LightGBM. In particular, we apply the following encoding strategies, including three standard approaches and two modified approaches: Angle, Amplitude, IQP, Entangled Angle, and Alternative IQP. As these approaches require either a significant number of qubits or gates to encode larger datasets, we perform feature selection to support the limited computing power of quantum simulators. This feature selection is performed through a quantum annealing enhanced approach that builds on a QUBO formulation of the problem. In this work, we provide a preliminary demonstration that quantum machine learning with the IQP encoding and LightGBM produce statistically equivalent results for a large majority of the assigned learning tasks.
\end{abstract}

\maketitle


\section{\label{sec:intro}Introduction}
As research into the viability and potential advantage of quantum machine learning continues, it is important to consider the effect that the data encoding has on the ability of the machine learning model to accurately learn from the target data. The typical approach to encoding data is a multi-qubit approach, where data is encoded through a circuit that projects the data into the Hilbert space, often through the use of rotation operations \cite{schuld2019quantum, havlivcek2019supervised}. This approach is conceptually similar to the idea of a kernel method, suggesting significant importance in the selection of the encoding method and its potential reliance on the structure of the underlying data \cite{schuld2021supervised}.

While the quantum equivalent of kernel methods are often applied to any quantum machine learning model, classical kernel methods are most often partnered with a support vector machine, where the goal is to maximize the distance between data points in different classes. This distance is typically maximized by projecting the data into a higher dimensional space that better separates the data and computing the distances in that space. In classical models, the kernel trick allows computation of these distances without needing to project the data into that higher dimensional space, instead using an alternate formulation that yields the inner product in that space \cite{hofmann2008kernel}. With quantum encoding, the data can be projected into the Hilbert space and the inner product computed directly in that space. Depending on the selection of encoding method, this projection can be performed efficiently with respect to the number of features. Similarly, the inner product can be efficiently computed through a measurement operation \cite{schuld2021machine}.

The primary goal of this work is to explore the effect different data encoding strategies have on several datasets that are well-suited for benchmarking at this scale. Although many datasets have significantly more features than can be feasibly simulated on classical hardware, this paper employs a quantum feature selection algorithm based on a quadratic unconstrained binary optimization (QUBO) formulation of the problem \cite{vlasic2022advantage}. As the goal of this work is to focus on the effect of the data encoding strategies, we restrict our selection of models to the quantum support vector classifier (QSVC), where the only quantum component is the data encoding. This approach facilitates a comparison with classical approaches and an analysis of the effects of quantum data encodings independent of a model with a variational quantum ansatz.

This work provides two novel contributions. First, this work provides a systematic analysis and comparison of major encoding strategies and classical approaches. Such an analysis is important to help shape future research and experiments in the field. Second, we demonstrate that for all of our selected datasets, the performance of the quantum IQP encoding and the LightGBM model are statistically equivalent. This equivalence suggests the potential for some benefit in using a quantum approach for machine learning.

The remainder of the paper is organized as follows. Section \ref{sec:exp} covers the parameters of our experiments, with Subsection \ref{subsec:data} covering the selected data, \ref{subsec:algos} the algorithms, and \ref{subsec:featmaps} the encoding feature maps. Section \ref{sec:results} discusses the results of the experiments and statistical analysis and Section \ref{sec:discussion} discusses those results and concludes the paper with a discussion on future work.

\section{\label{sec:exp}Experiments}

\subsection{\label{subsec:data}Data}
The selection of the data had five criteria: a binary classification problem, no missing values, small difference between the number in each class, heterogeneity between each of the data sets, and selecting from respected open repositories. From the criteria, the UCI Machine Learning Repository was selected, and from the repository the binary classification datasets of Ionosphere (Ionosphere), Connectionist Bench (Sonar), Sirtuin 6 Small Molecules (Sirtuin6), and Breast Cancer Wisconsin (WDBC)
\cite{misc_ionosphere_52,misc_connectionist_bench_151,misc_sirtuin6_small_molecules_748,misc_breast_cancer_wisconsin_17} were identified. 

The dimension of the features for three out the four data sets are were too large for the computation. Adjusting for this shortcoming, a feature selection algorithm \cite{mucke2023feature}, enhanced by quantum annealing, is applied to down select; this algorithm is described in Section \ref{subsec:algos}. The description of each data set follows.

\begin{figure}[htbp]
    \begin{adjustbox}{valign=t}
    \subfigure[\label{subfig:corr-ionoshpere}Pearson correlation between each feature with the Ionosphere dataset.
    ]{ \includegraphics[scale=.05]{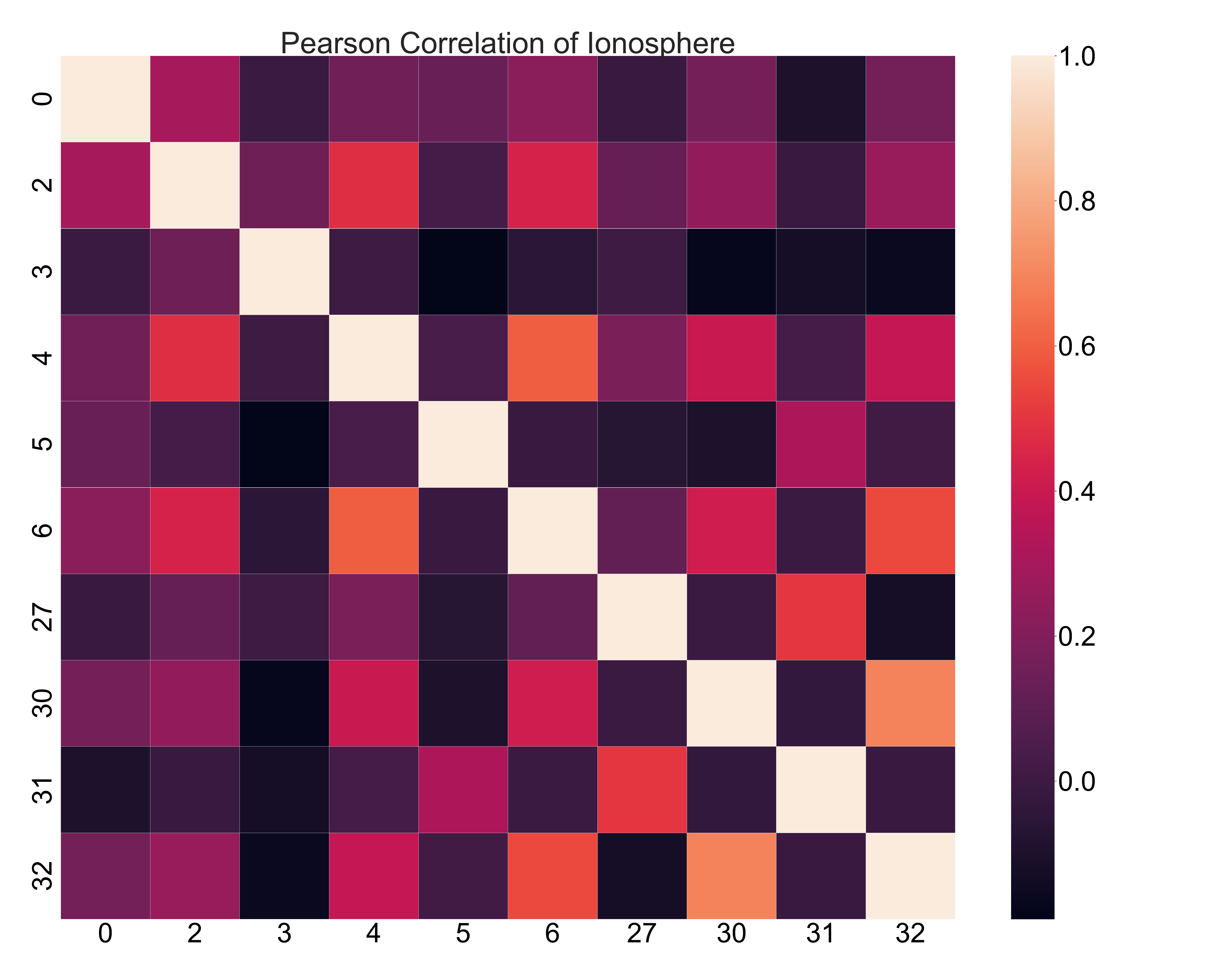}
    }
    \end{adjustbox}
    \begin{adjustbox}{valign=t} 
    \subfigure[\label{subfig:corr-wdbc}Pearson correlation between each feature with the WDBC dataset.]{\includegraphics[scale=.05]{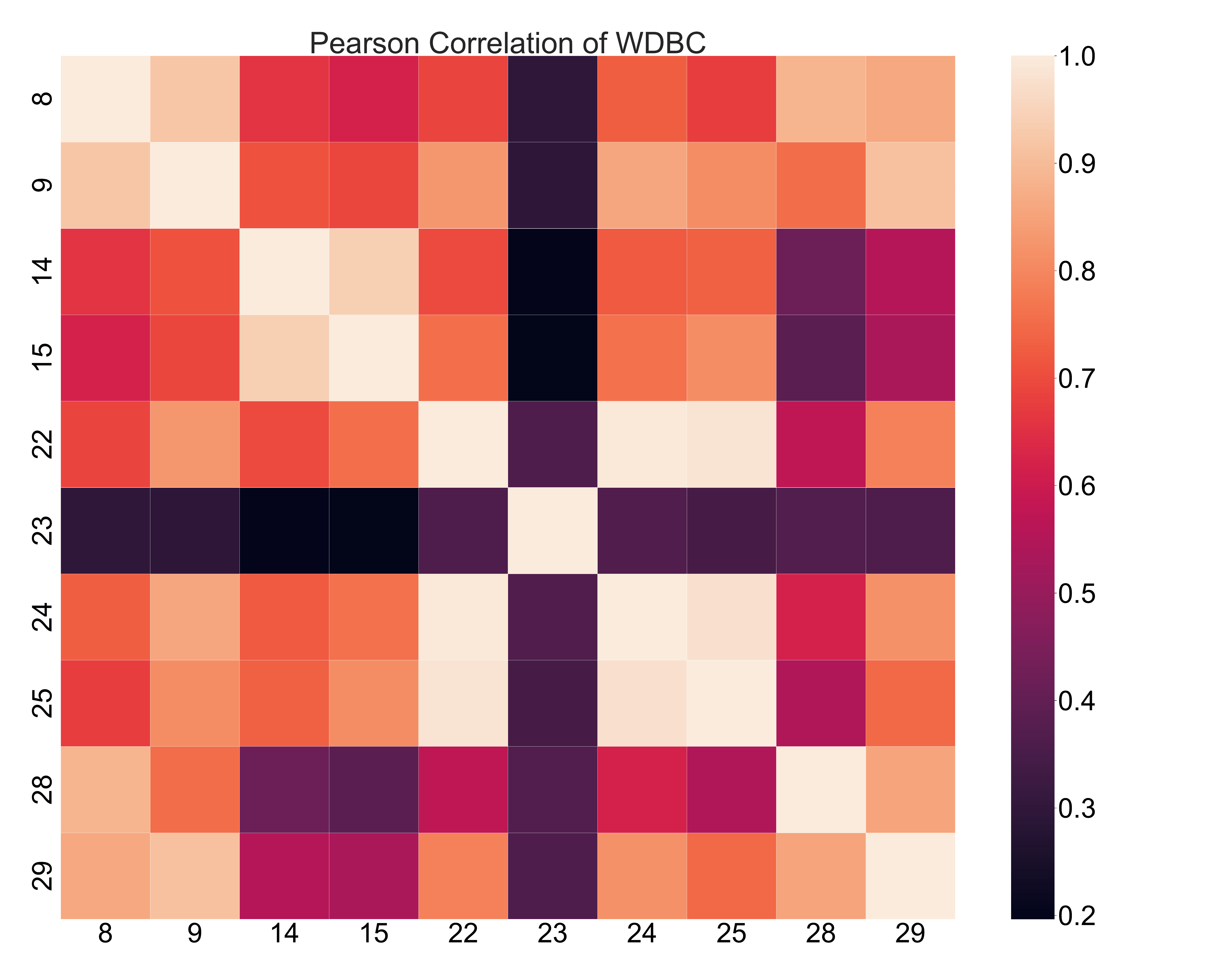} 
    }
    \end{adjustbox}
    \begin{adjustbox}{valign=t} 
    \subfigure[\label{subfig:corr-sirt6}Pearson correlation between each feature with the Sirtuin6 dataset.]{ \includegraphics[scale=.05]{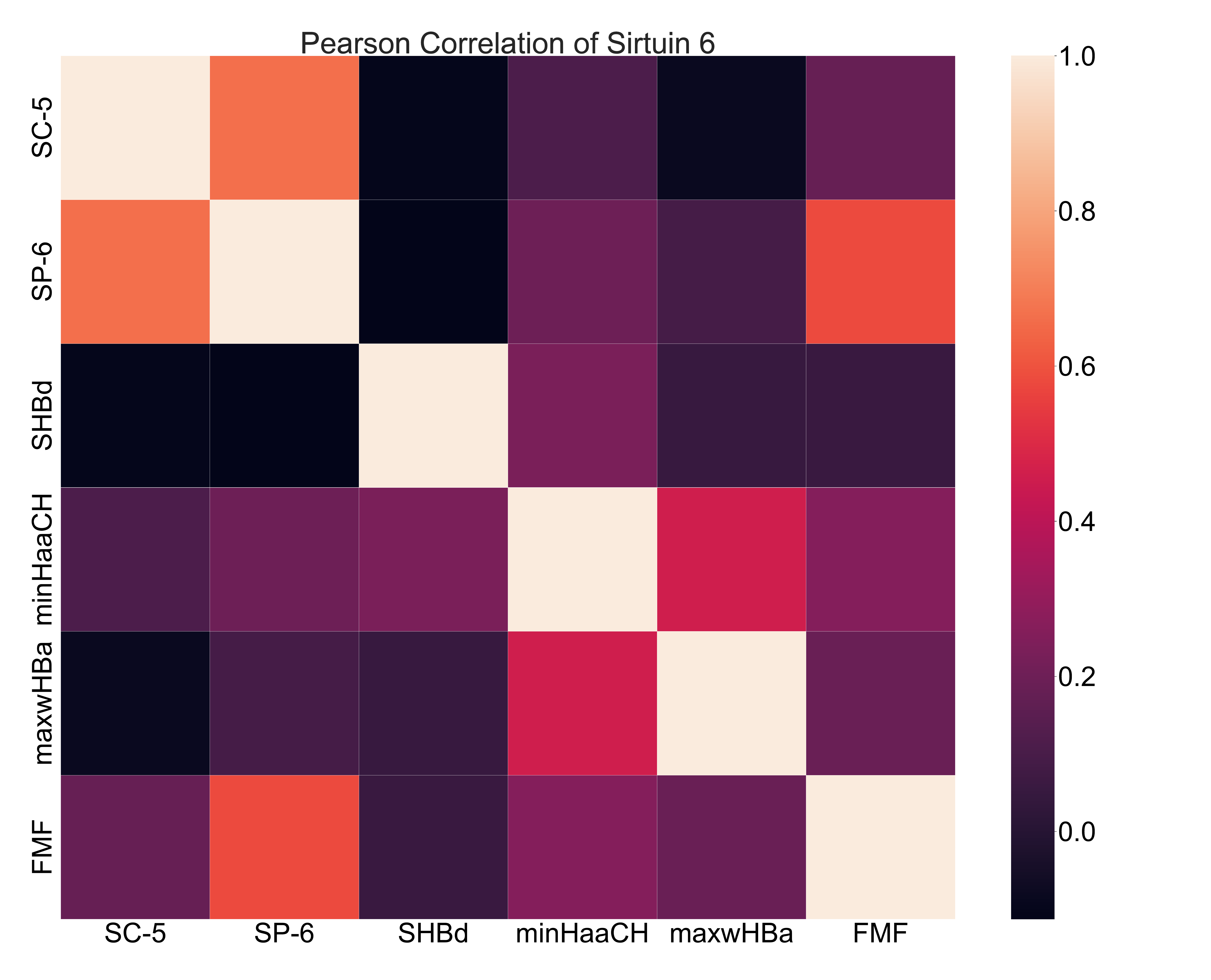}
    }
    \end{adjustbox}
    \begin{adjustbox}{valign=t} 
    \subfigure[\label{subfig:corr-sonar}Pearson correlation between each feature with the Sonar dataset.]{ \includegraphics[scale=.05]{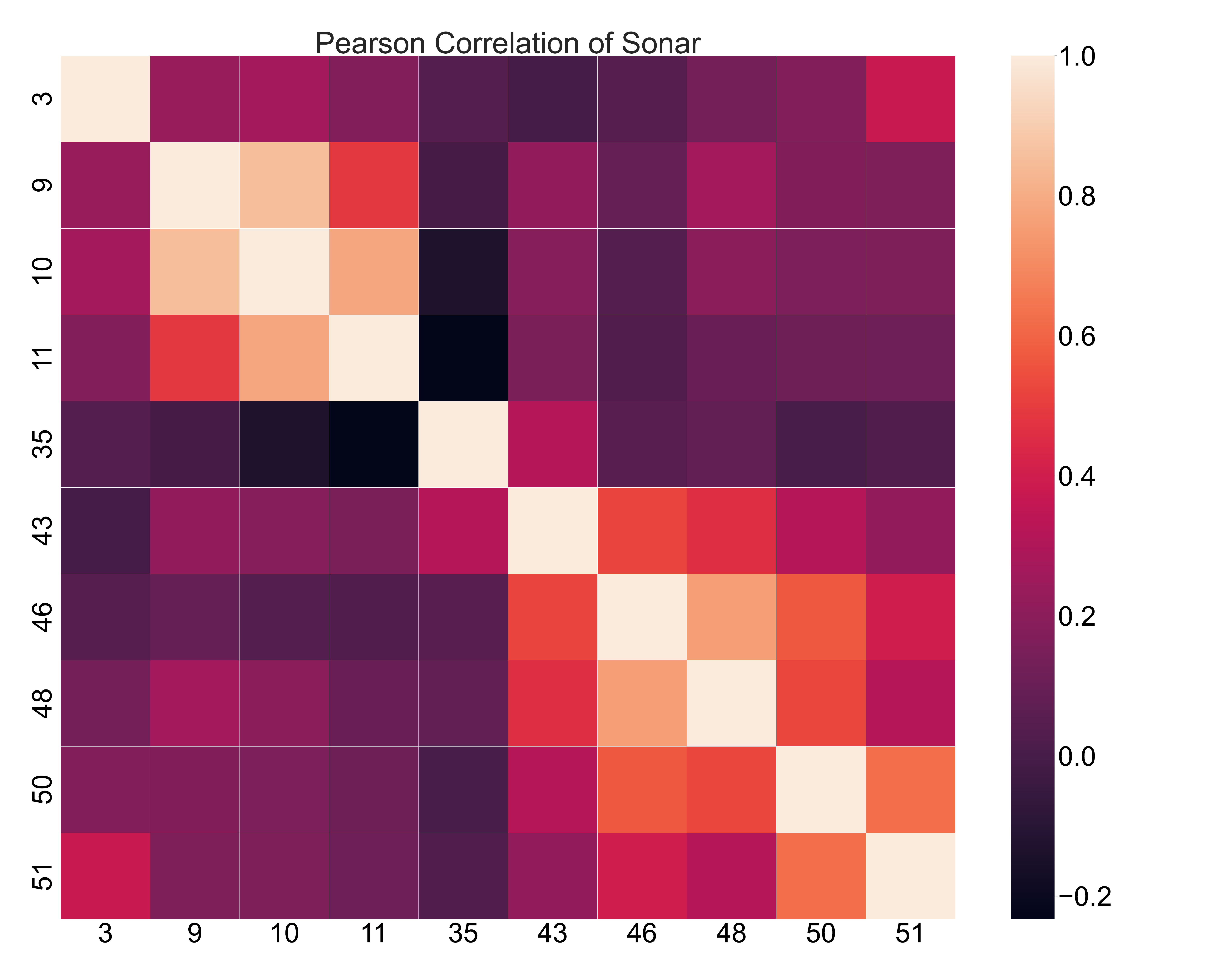}
    }
    \end{adjustbox}
\caption{\label{fig:corrs}Pearson correlation scores of the features for each respective dataset.}
\end{figure}

For the Ionosphere dataset (Ionosphere) \cite{misc_ionosphere_52} there is the task of identifying radar that maps the ionosphere (good), and radar where the signals pass through the ionosphere (bad). There are 351 data points, 33 features that are all continuous, and a ratio of 1.79:1 between the two classes. The down selected columns  are  `0',  `2',  `3',  `4',  `5',  `6', `27', `30', `31', and `32'. Figure \ref{subfig:corr-ionoshpere} displays the Pearson correlation between the features, showing a fairly intricate dataset. 

The Connectionist Bench data (Sonar) \cite{misc_connectionist_bench_151} consists of identifying whether a piece of ground is a mine, or metal cylinder, or just a rock after sonar is applied at various angles. The data set has 208 data points, 60 continuous features, and a ratio of 1.14:1 between the two classes. The columns selected are `3',  `9', `10', `11', `35', `43', `46', `48', `50', `51'. Figure \ref{subfig:corr-sonar} displays the Pearson correlation between the features. 

The Sirtuin6 Small Molecules (Sirtuin6) \cite{misc_sirtuin6_small_molecules_748} is tasked with identifying which small molecules are candidate inhibitors of the target protein. The data contains 100 data points, 6 continuous features, and a ratio of 1:1 between the two classes. Features were calculated and selected a priori by the authors. Figure \ref{subfig:corr-sirt6} displays the Pearson correlation between the features. 

The Breast Cancer Wisconsin (WDBC) \cite{misc_breast_cancer_wisconsin_17} has the clear problem of identifying cancer. The data has 569 data points, 30 continuous features that were derived from images, and a ratio of 1.68:1 between the two classes. The columns selected are `8', `9', `14', `15', `22', `23', `24', `25', `28', `29'. Figure \ref{subfig:corr-wdbc} displays the Pearson correlation between the features. 

\subsection{\label{subsec:algos}Algorithms}

The QSVM is constructed by building on a classical SVM. With a SVM, the dividing hyperplane is found by computing the distances between pairs of points. In a classical SVM, this distance can be computed with respect to a kernel through the kernel trick \cite{crammer2001algorithmic}. In the quantum analog, this distance can be directly computed in the quantum Hilbert space. To do so, the two data points ($x_1$ and $x_2$) may be embedded and compared using the SWAP test. However, the comparison can be done with half as many qubits by performing the embedding of $x_1$ followed by the inverse of the embedding of $x_2$. By projecting this state onto the initial state $\ket{0\ldots0}\bra{0\ldots0}$ and measuring the result, the distance between the two points is given \cite{schuld2021supervised}. 

Once computed in the quantum space, this distance can be passed to a classical SVM, enabling separation of the data within a quantum space and definition of the hyperplane using the classical SVM algorithm. Qiskit's QSVM implementation is particularly flexible, as it offers both binary and multiclass classification, the choice of which is automatically performed internally \cite{aleksandrowicz2019qiskit}.

Computational shortcomings, except for the Sirtuin6 data set, did not allow the use of all the columns for the data sets. Given the loss of information through traditional dimension reduction methods, such as principal component analysis, feature selection was selected. However, when selecting features there has to be a balance between the signal to the target for a feature and the shared signals between other features, ergo, balancing bias and variance, ensuring collective information is maintained during the down selection process. 

For this task applied the feature selection algorithm derived by M{\"u}cke et al. \cite{mucke2023feature}, and was shown by Vlasic et al. \cite{vlasic2022advantage} to have an advantage when down selecting features. The algorithm uses the quadratic binary unconstrained optimization (QUBO) formulation where the decision variables are the features, the linear terms is the correlation of the feature and target variable, and the quadratic terms. The linear terms are weighted with a constant in the unit interval, and the quadratic terms are weighted with the compliment of this constant. With a given number of features, the binary search of a sorted array algorithm is utilized to calculate which weight will yield that number of features.

Havl{\'\i}{\v{c}}ek et al. \cite{havlivcek2019supervised} translated support vector machines \cite{pradhan2012support} into quantum computation. 
QSVC is implemented in Qiskit \cite{aleksandrowicz2019qiskit} and the feature selection algorithm is implemented as proposed.

\subsection{\label{subsec:featmaps}Feature Maps}
For the experiments, five feature maps were selected. Three of the maps are standard, Angle, Amplitude, and IQP, while the other two are an extension of Angle and IQP, Entangled Angle (EntAngle) and an alternative formulation of an IQP circuit (AltIQP).

Angle encoding \cite{schuld2019quantum,skolik2021layerwise} is an intuitive map where the entries of the data point, $d$, are fed into the circuit as the angles of gates on the respective wire. This method has been noted to not fully leverage quantum since gates that create entanglement are not utilized. There is an analytic form of the map, where for $X_{i}$ the Pauli $X$ gate acting on the $i^{th}$ qubit and $d_i$ is the $i^{th}$ entry of the data point, the encoded operator has the mapped form
\begin{equation}
     d \longrightarrow \bigotimes_{l=1,2, \ldots,n} \exp( -i X_l d_l ) \ket{0\ldots0}.
\end{equation}

\begin{figure}[htbp]
   \begin{adjustbox}{valign=t}
    \begin{quantikz}[thin lines] 
        \lstick{$\ket{0}$} & \gate{H} & \gate{R_y(\theta_1 )}  & \ctrl{1} & \qw & \targ{} & \qw
       \\ \lstick{$\ket{0}$} & \gate{H} & \gate{R_y(\theta_2 )} & \targ{} & \ctrl{1} & \qw & \qw
       \\ \lstick{$\ket{0}$} & \gate{H} & \gate{R_y(\theta_3 )} & \qw & \targ{} & \ctrl{-2} & \qw
    \end{quantikz}

\end{adjustbox}
    \caption{\label{fig:alt_ang}Two-qubit Entangled Angle example}
\end{figure}
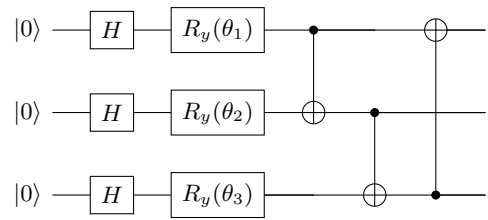

The Entangled Angle encoding seeks to build on the Angle encoding strategy by further leveraging quantum mechanics through the use of entanglement. This circuit is constructed by applying a Hadamard gate to each wire, followed by the typical angle encoding, and a Controlled-NOT gate between every adjacent pair of wires. A three-qubit version of the circuit is shown in Figure \ref{fig:alt_ang}. Of particular importance is the use of the Pauli $Y$ gate for the angle embedding as the state \ket{+} is not measurably affected by Pauli $X$ or $Z$ rotations.

\begin{figure}[htbp]
   \begin{adjustbox}{valign=t}
    \subfigure[ Binary Tree Decomposition]{ \begin{forest}
      for tree={l+=.25cm} 
      [$1$
        [$\sqrt{.6}$[$\sqrt{.2}$][ $\sqrt{.4}$ ]]
        [$\sqrt{.4}$[$\sqrt{.3}$][$\sqrt{.1}$]]
      ]
    \end{forest} }
    \end{adjustbox}
    \begin{adjustbox}{valign=t} 
    \subfigure[ Encoding of the Binary Tree]{ \begin{quantikz}[thin lines] 
        \lstick{$\ket{0}$} & \gate{R_y\Big(\sqrt{.6} \Big)}  & \octrl{1} &\ctrl{1}& \qw
       \\ \lstick{$\ket{0}$} & \qw & \gate{R_y\Big(\sqrt{.3/.4 }\Big)}&  \gate{R_y\Big(\sqrt{ .2/.6}\Big)} & \qw
    \end{quantikz} }
\end{adjustbox}
\caption{\label{fig:amp_enc} This is an example on how to amplitude encode data: (a) displays the binary tree decomposition of $( \sqrt{.2}, \sqrt{.4}, \sqrt{.3}, \sqrt{.1})$, and (b) is the respective circuit to encode the binary tree.}
\end{figure}
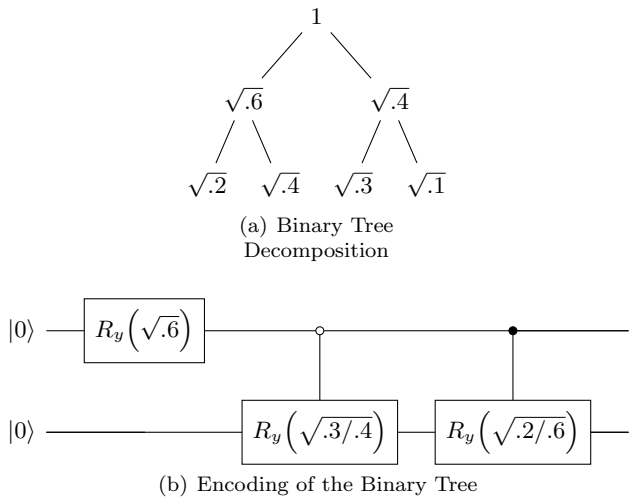

The Amplitude encoding \cite{grover2002creating,araujo2021divide} takes a data point $d$ of the form $\displaystyle \sum_{i=1}^{n} |d_i|^2 = 1$, with $n=2^m$ for $m\in\mathbb{N}$, into a binary tree structure where the partitions are data points split in half and built back by leveraging Bayes. To illuminate Amplitude consider the simple example of the data point $( \sqrt{.2}, \sqrt{.4}, \sqrt{.3}, \sqrt{.1})$. This data point will be decomposed through Amplitude encoding in order to create the state $\sqrt{.2} \ket{00} + \sqrt{.4} \ket{10} + \sqrt{.3} \ket{01} + \sqrt{.1} \ket{11}$. The binary tree and subsequent circuit is given in Figure \ref{fig:amp_enc}.
The authors in \cite{araujo2021divide} derive another technique, which they denote as divide-and-conquer, to encode the amplitude of a data point. It this this method that was utilized in the experiments. Given the intricacy of the algorithm the description will be left out, and the reader is encouraged to read the paper.

The IQP mapping technique \cite{havlivcek2019supervised, bremner2016average} takes its ansatz from physics and transforms data points into a second order Hamiltonian. However, this order can be extended. The authors assume the data point $d$ is in the space $(0,2\pi]^n$ and, denoting $Z_i$ as the Pauli $Z$ gate acting on the $i^{th}$ qubit, define the unitary operator \begin{equation}\label{used}
U_Z(d) = \exp\left( \sum_{i=1}^n d_i Z_i + \sum_{i=1}^n \sum_{j=1}^n (\pi - d_i)(\pi- d_j) Z_i Z_j \right).
\end{equation} The general form of this encoding method is $\displaystyle U_{\Phi}(x) = \exp\left( \sum_{S \subset \{1,2,\ldots,n\}} \Phi_{S}(x) \prod_{i\in S}Z_i \right)$. The coefficients in the quadratic terms are centered around $0$ with standard deviation of $1$. Denoting $H$ as the Hadamard gate, the IQP map is defined as 
\begin{equation}
d \rightarrow U_Z( d) H^{\otimes n} U_Z( d ) H^{\otimes n} \ket{0\ldots0}.
\end{equation}
This is not the only IQP map, in fact, the authors in \cite{li2022quantum} derived a similar encoding technique by considering Ising interactions of the unitary operators in $U_Z$, and where the Hadamard gates add uniform superpositions. The method titled IQP within this paper is derived directly from \cite{bremner2016average} whereas the Alternative IQP method is the implementation proposed in \cite{havlivcek2019supervised}.

\section{\label{sec:results}Results}
The experiments consist of training 50 models per data set and per feature map, where the 50 models are derived by splitting the test/train data subsets 50 times. While more model would considerably decrease standard error, the standard error and mean from 50 models is sufficient enough to average out outlier scores. However, the time to train a model is the reason 50 model was chosen.


\begin{figure*}[!ht]
\begin{center}
\begin{tabular}{cccc}
\subf{\includegraphics[width=45mm]{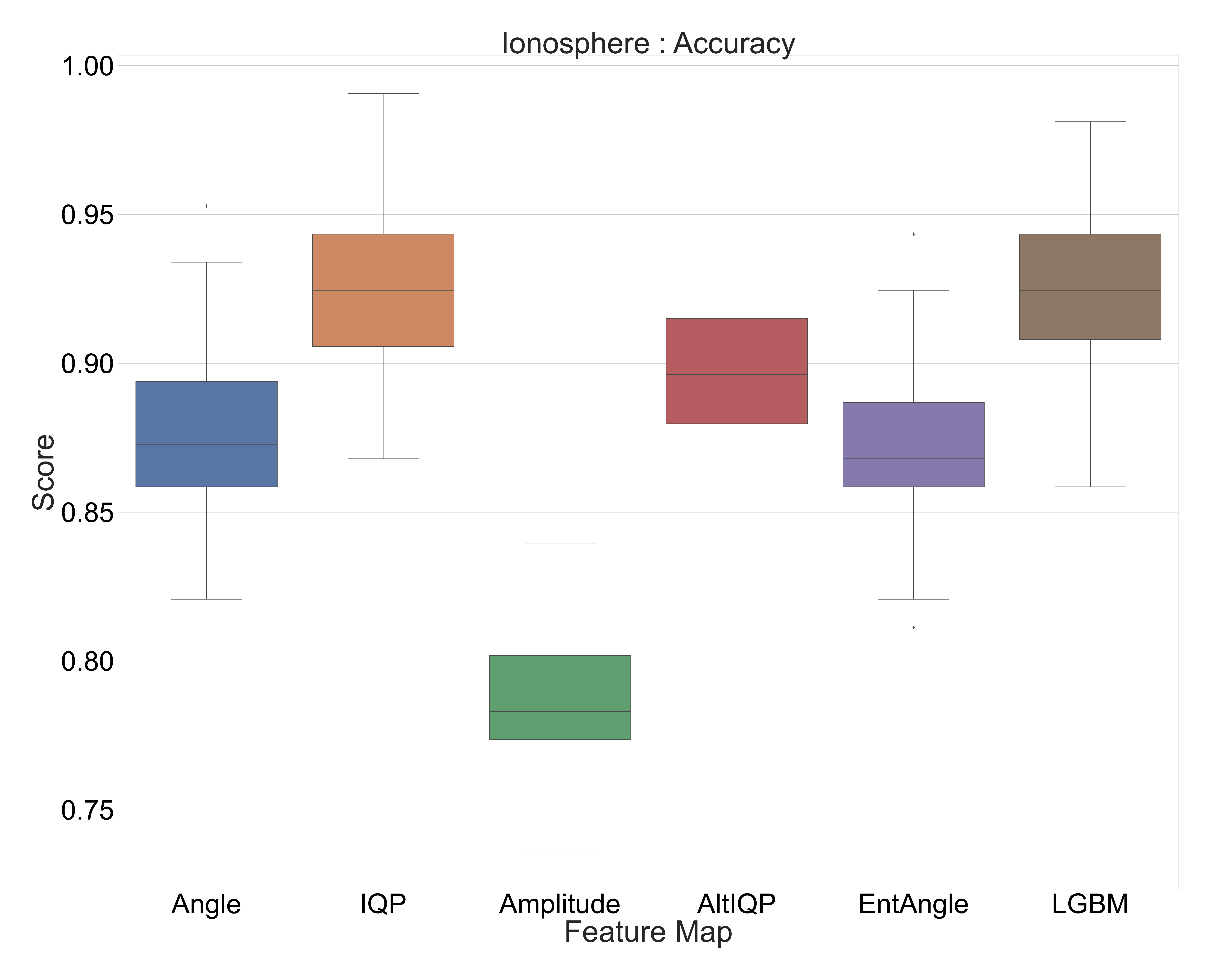}}
     {Ionosphere Accuracy}
&
\subf{\includegraphics[width=45mm]{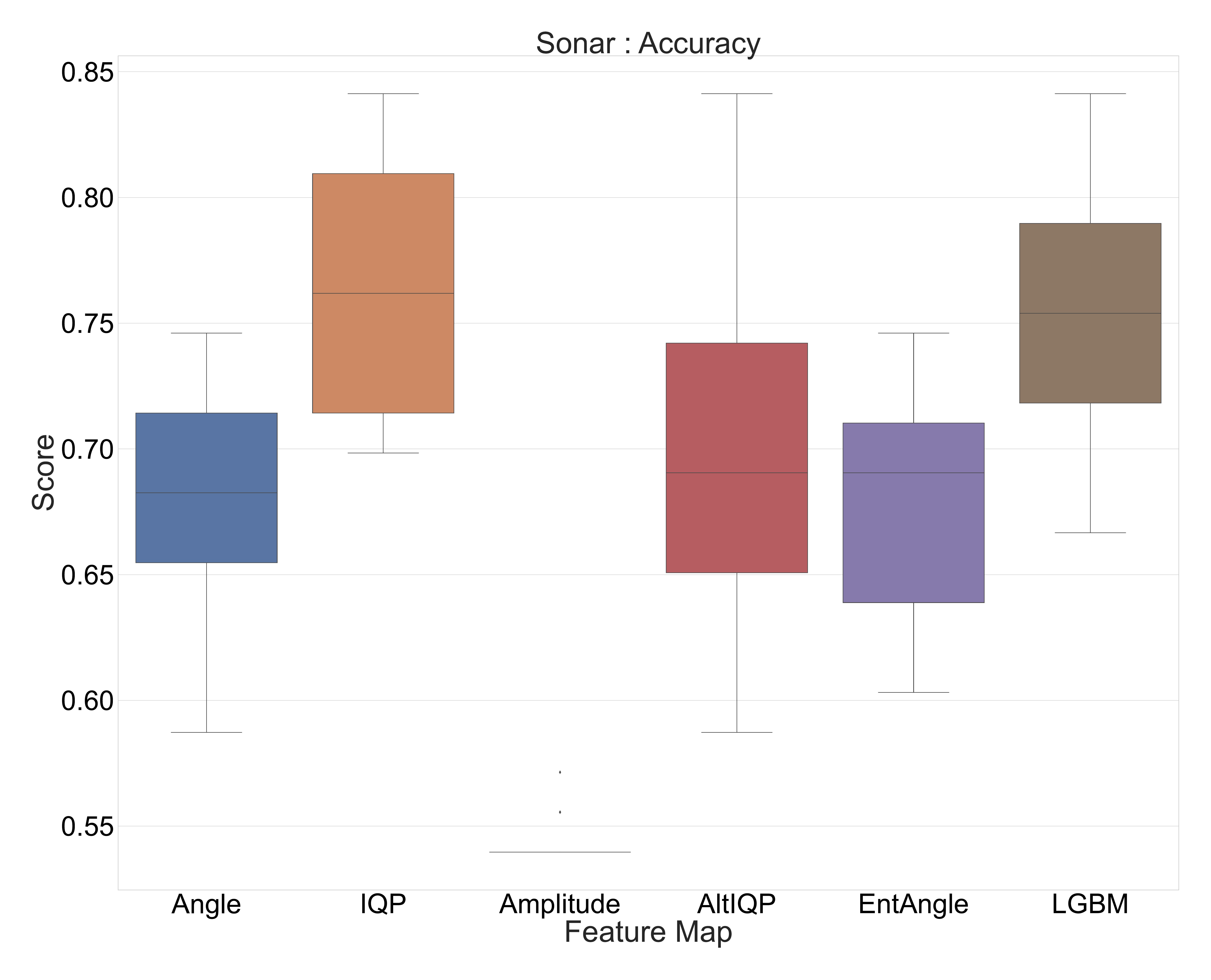}}
     {Sonar Accuracy}
&
\subf{\includegraphics[width=45mm]{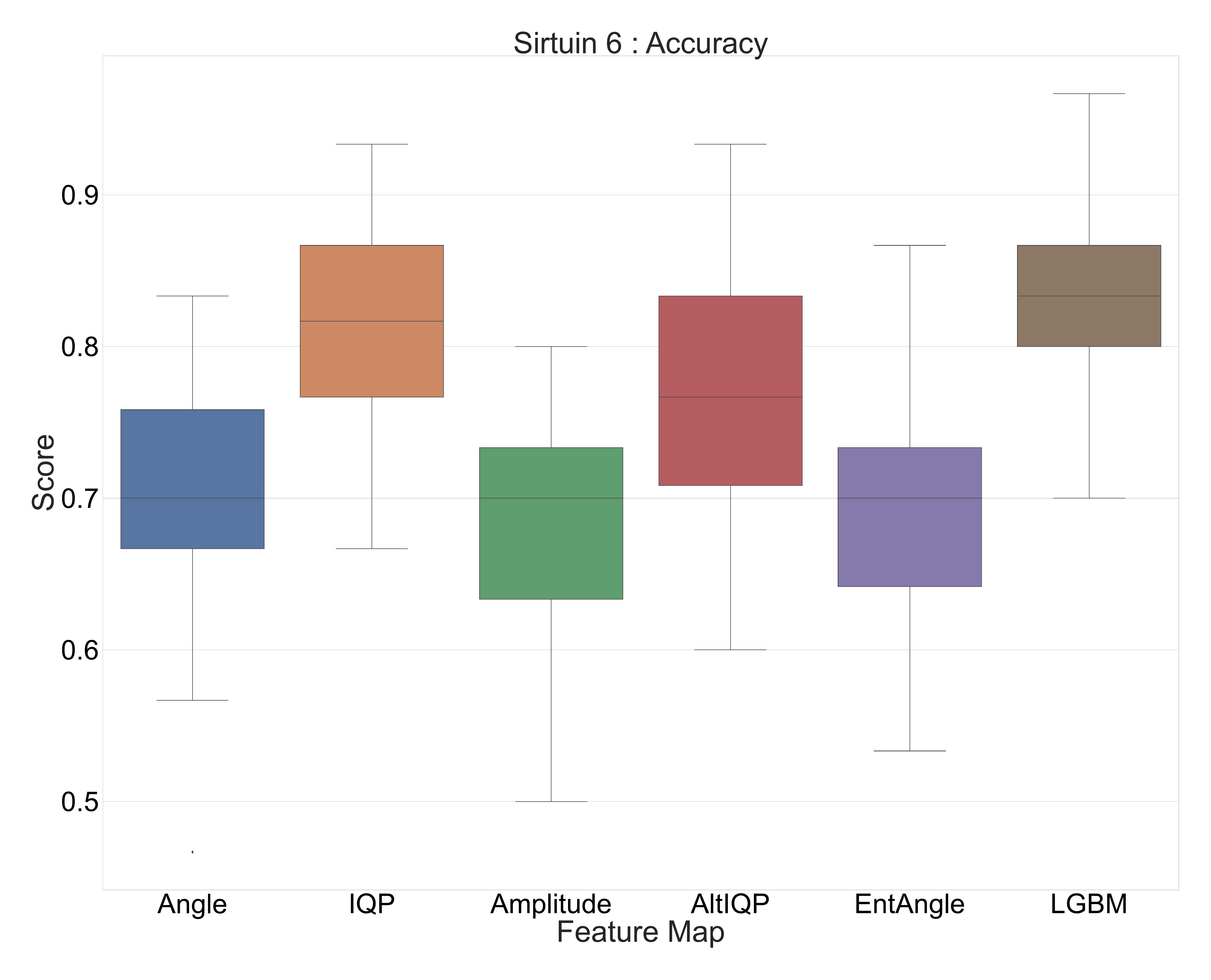}}
     {Sirtuin6 Accuracy}
&
\subf{\includegraphics[width=45mm]{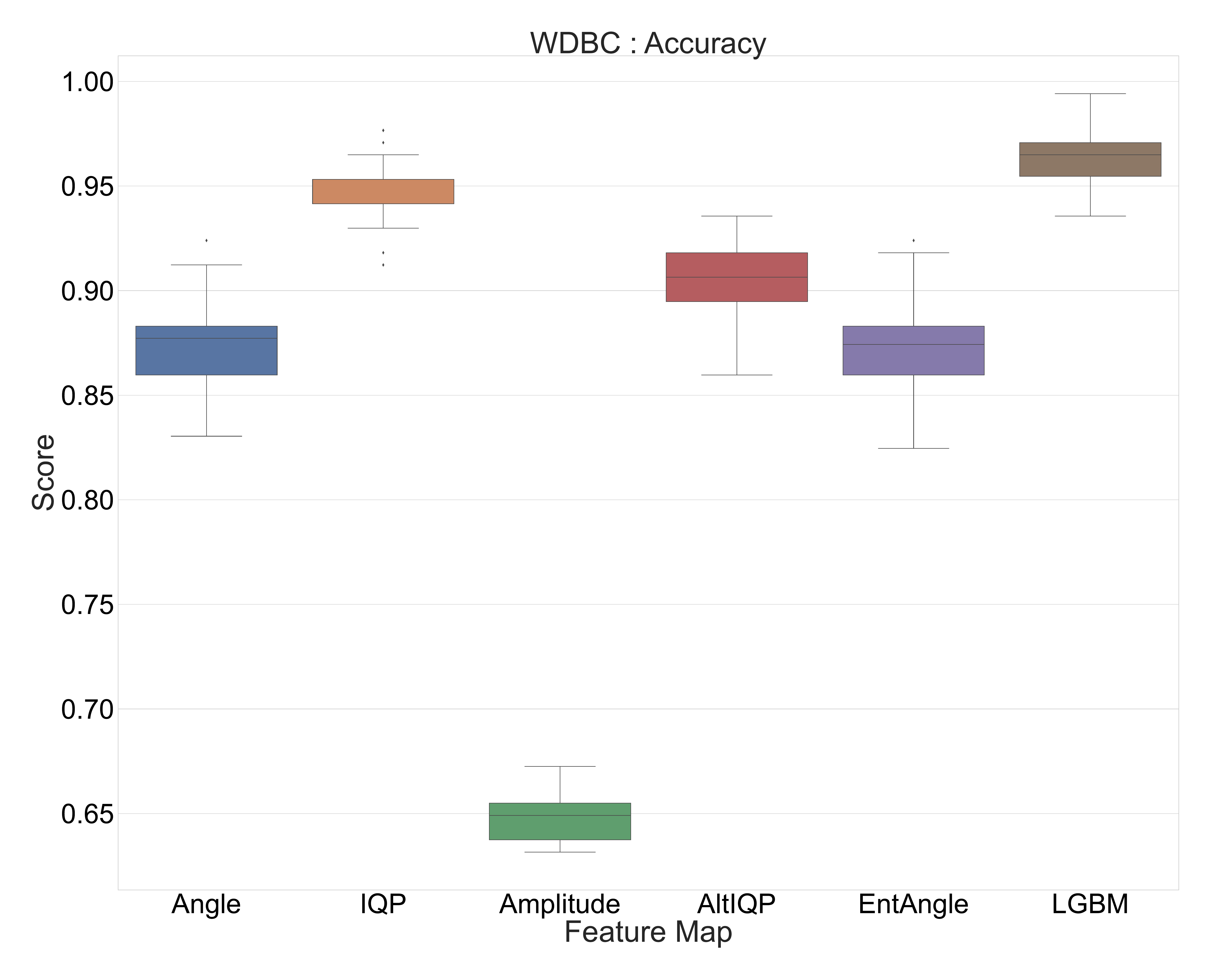}}
     {WDBC Accuracy}
\\
\subf{\includegraphics[width=45mm]{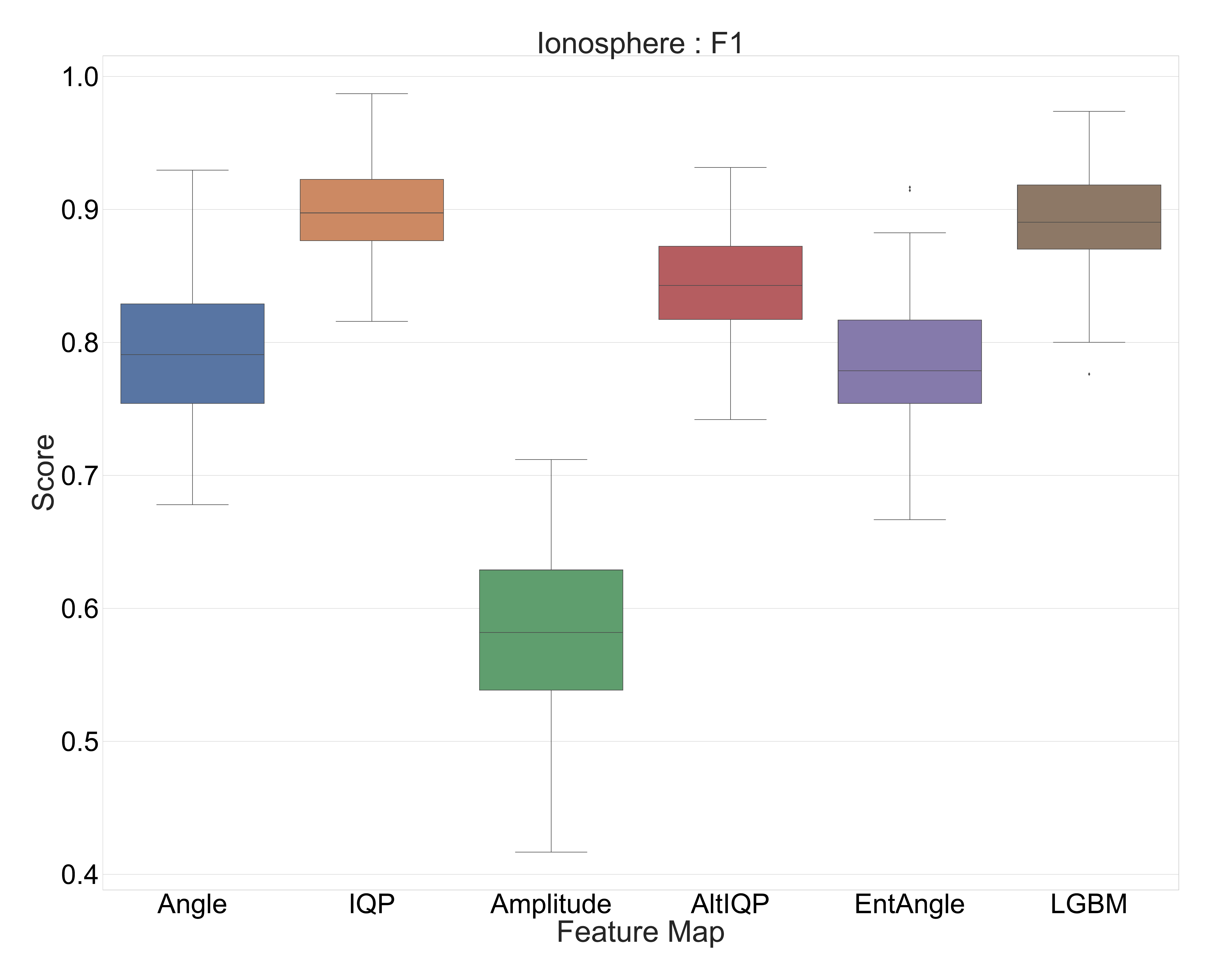}}
     {Ionosphere F1}
&
\subf{\includegraphics[width=45mm]{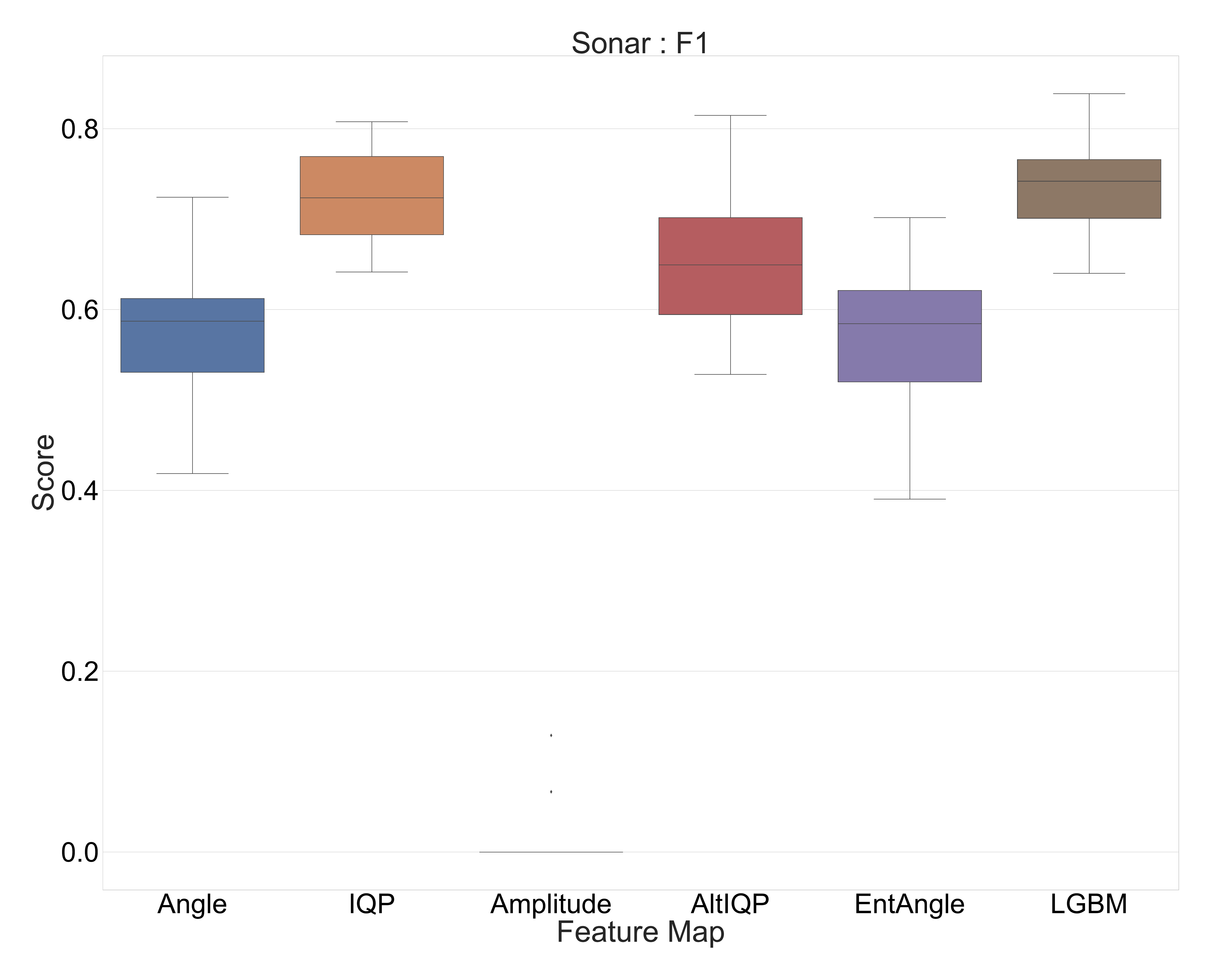}}
     {Sonar F1}
&
\subf{\includegraphics[width=45mm]{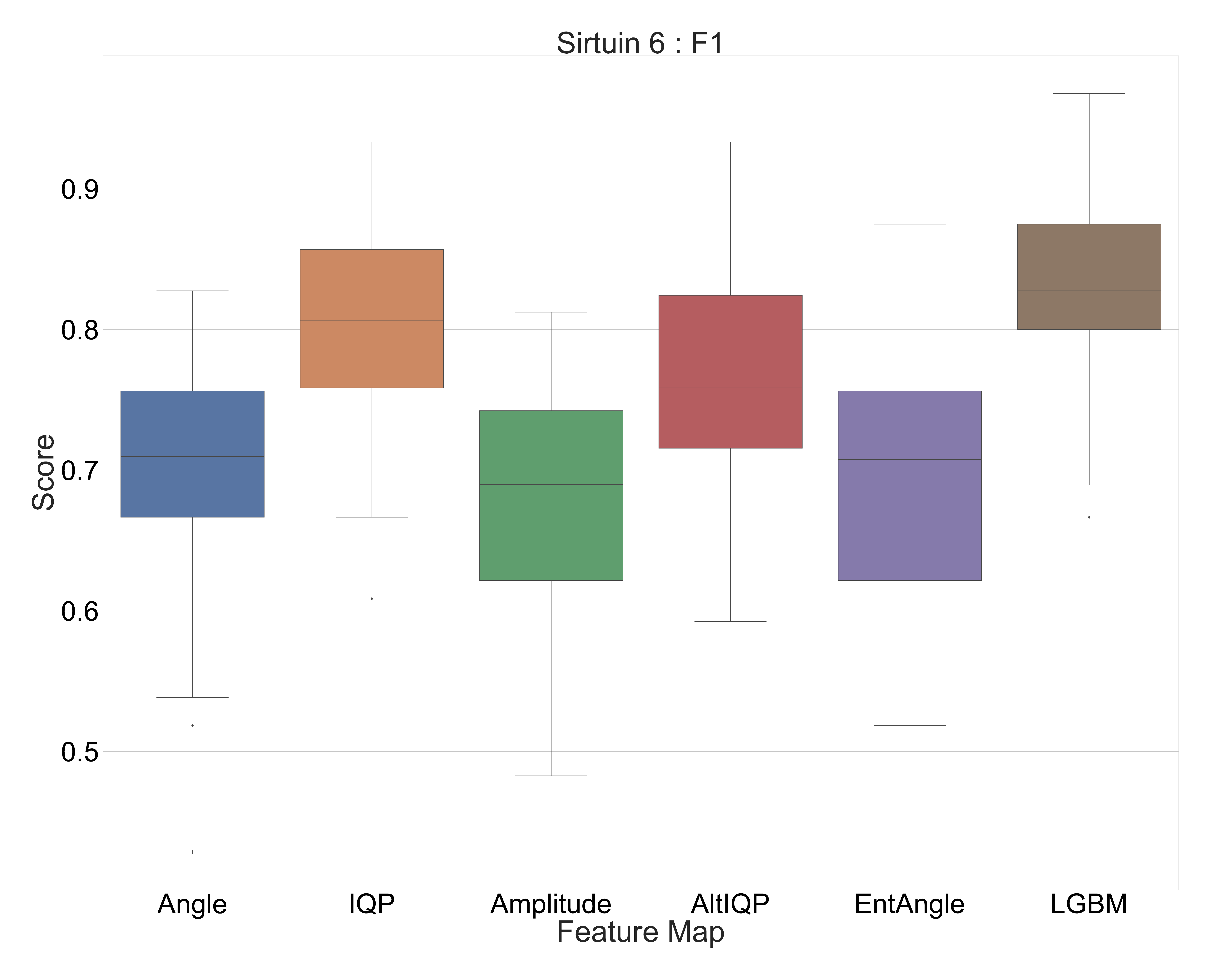}}
     {Sirtuin6 F1}
&
\subf{\includegraphics[width=45mm]{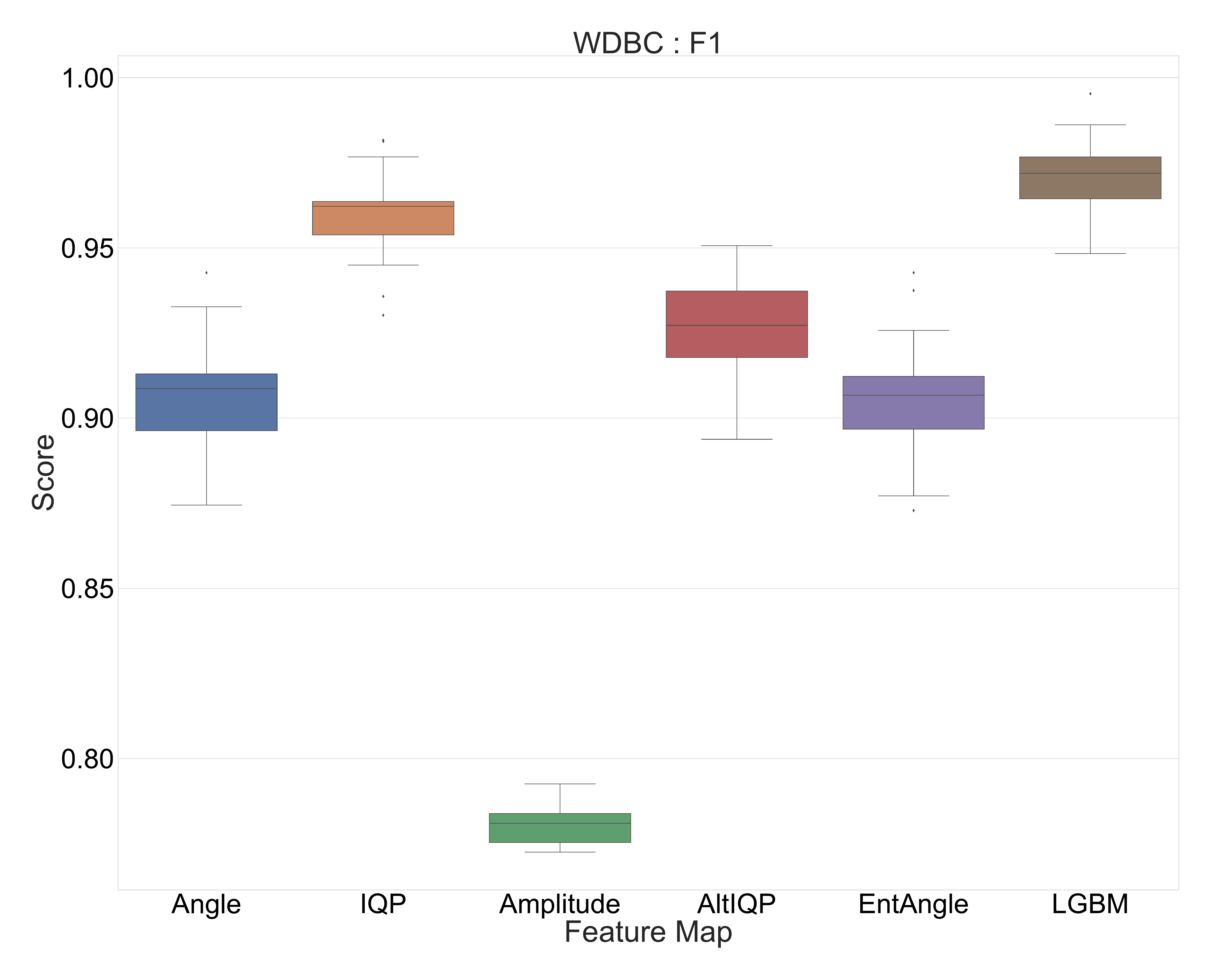}}
     {WDBC F1}
\\
\subf{\includegraphics[width=45mm]{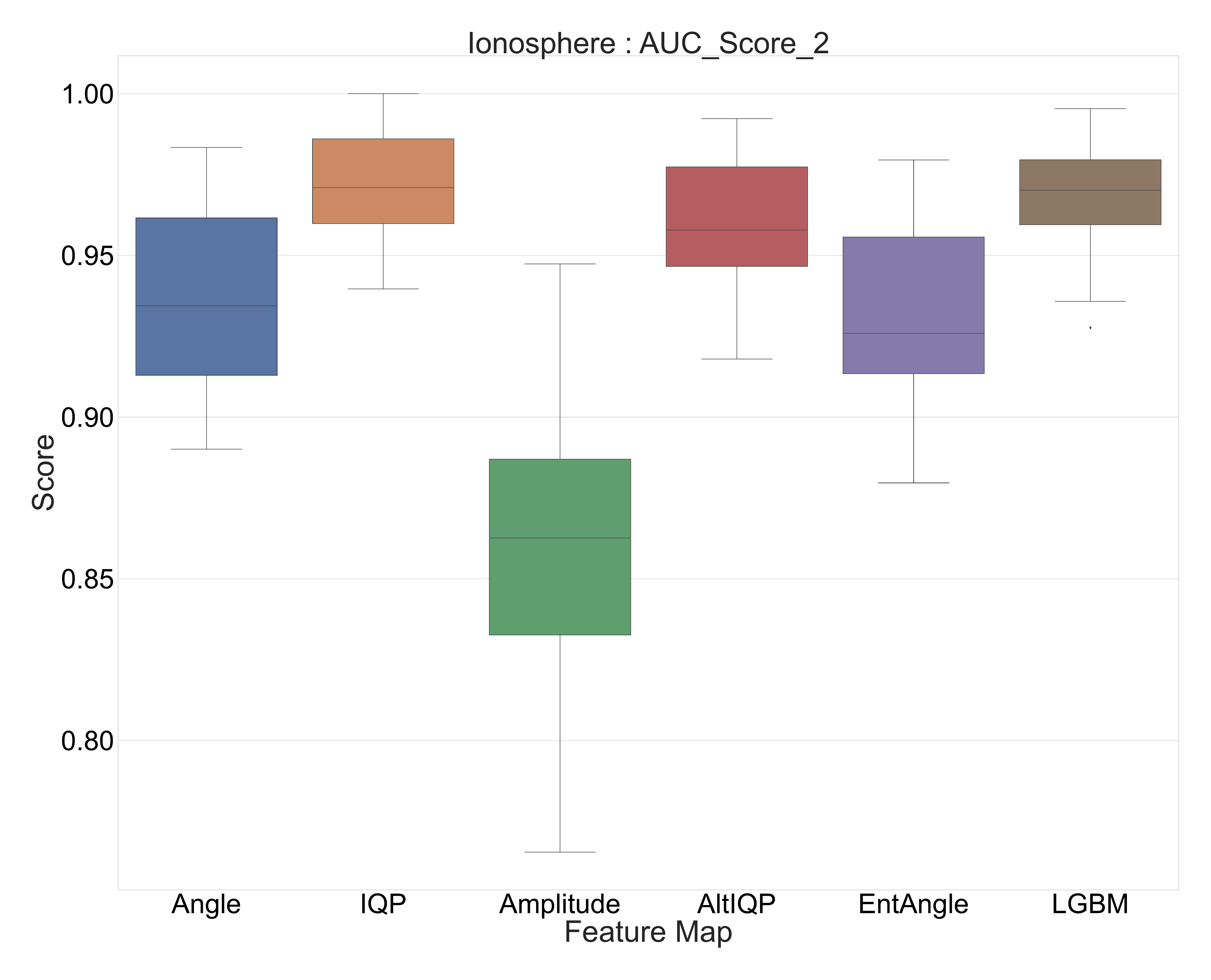}}
     {Ionosphere AUC}
&
\subf{\includegraphics[width=45mm]{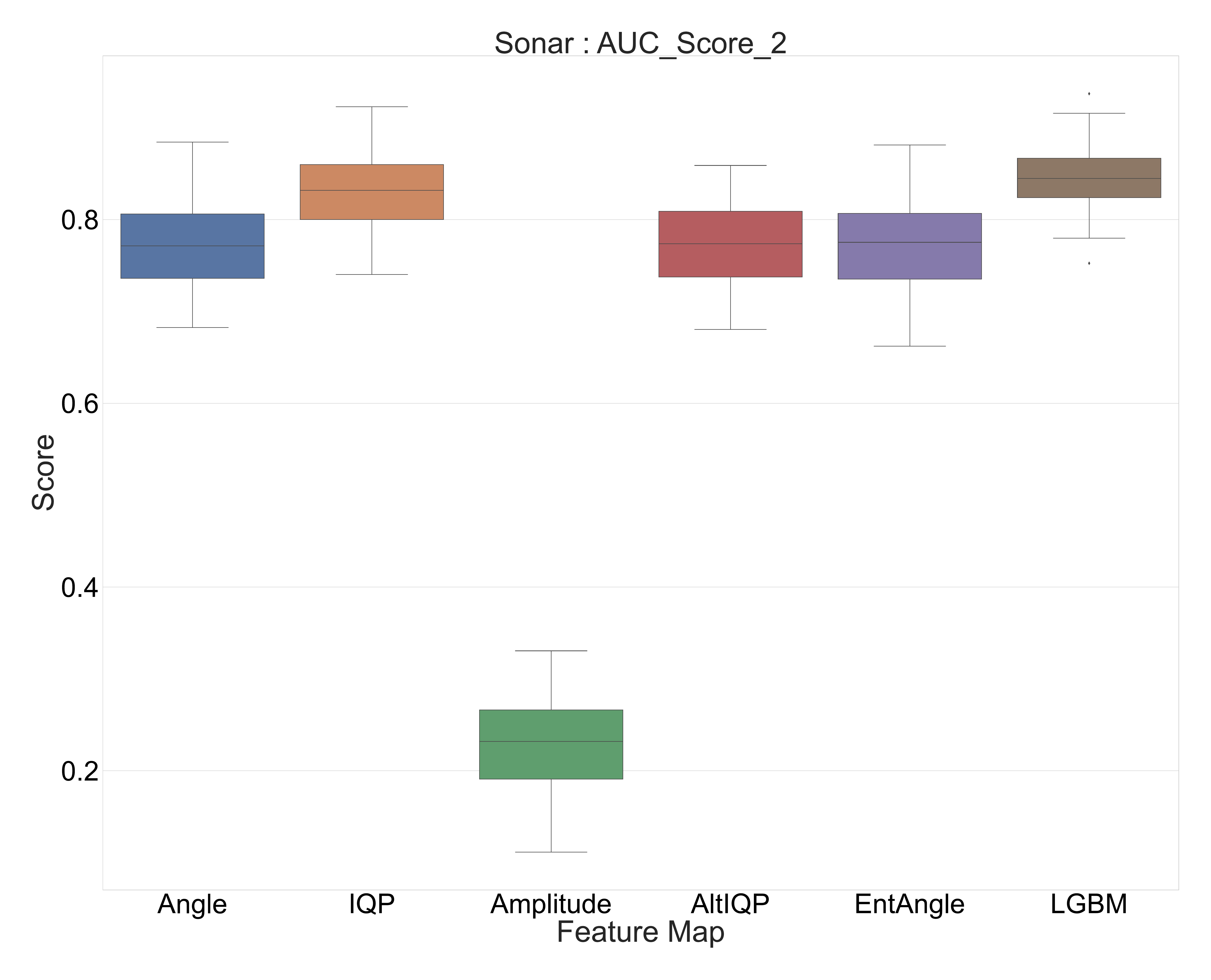}}
     {Sonar AUC}
&
\subf{\includegraphics[width=45mm]{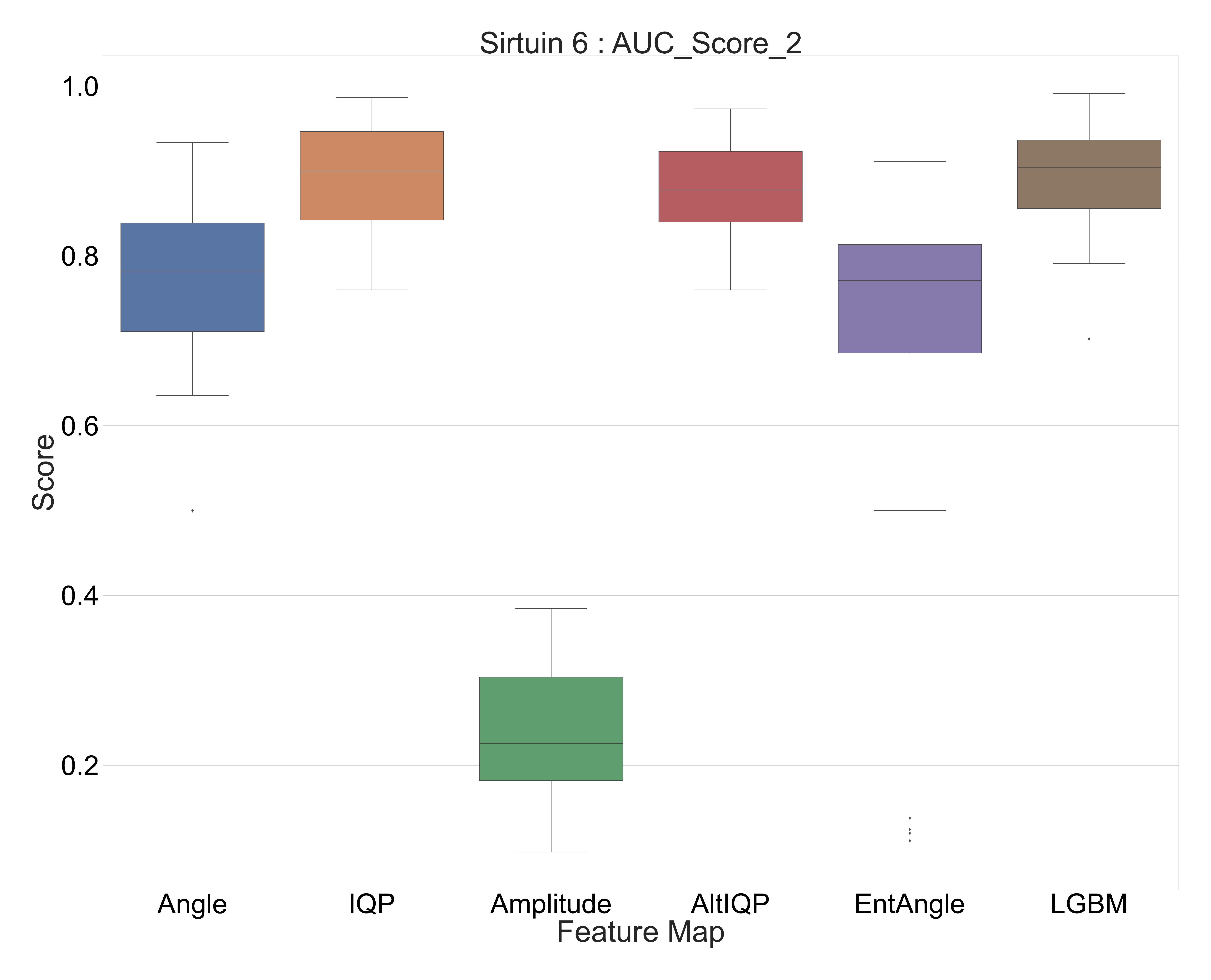}}
     {Sirtuin6 AUC}
&
\subf{\includegraphics[width=45mm]{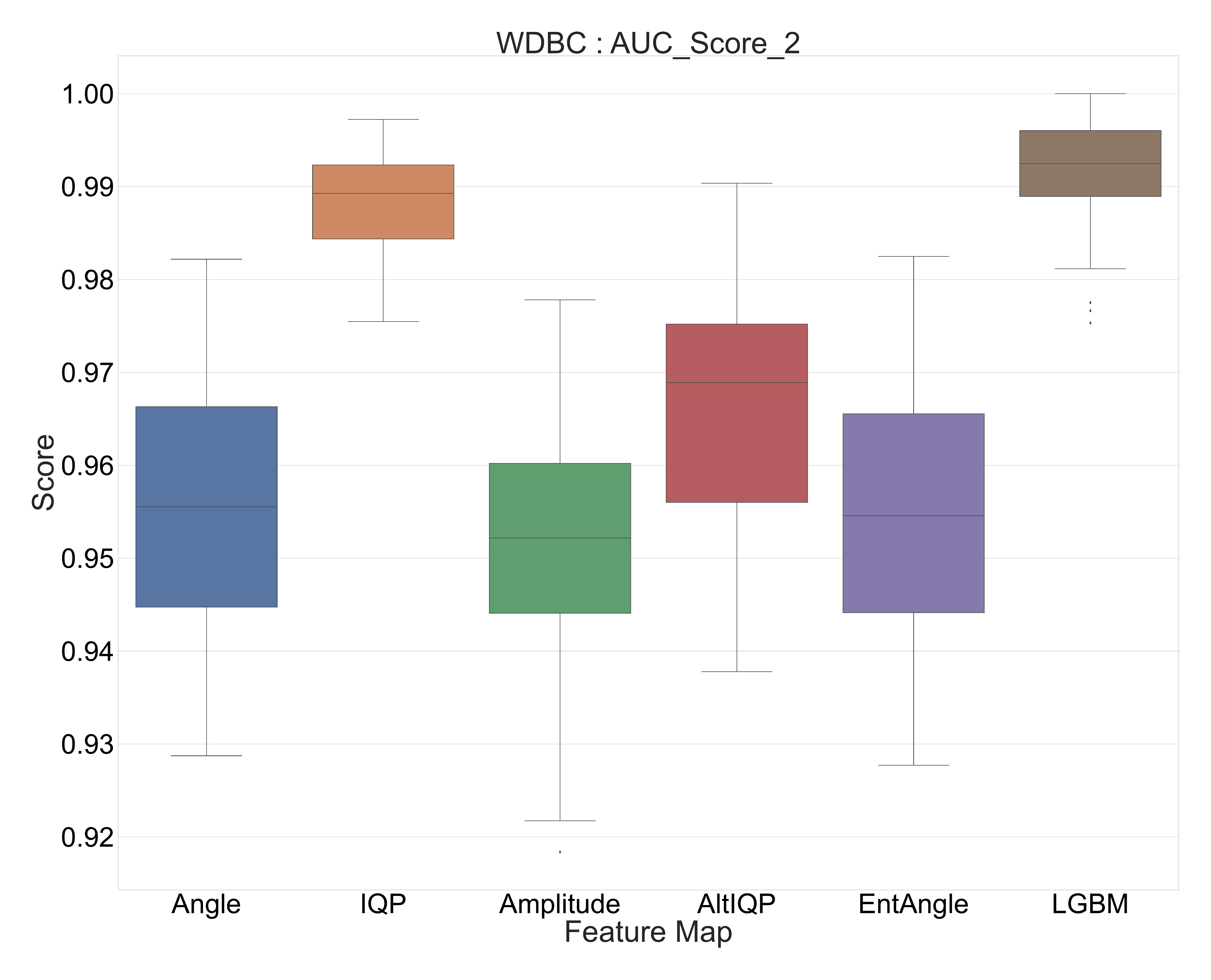}}
     {WDBC AUC} \\
\end{tabular}
\caption{\label{fig:results}The aggregated results of the experiments are collected. Each column consists of a data set with the accuracy, F1 score, and AUC score described in box plots.}
\end{center}
\end{figure*}

\twocolumngrid

For consistency, a random seed is selected to yield 50 random integers, which are then used to seed the test/train split. Every test/train split was $70\% \big/ 30\%$; the Scikit Learn package \cite{pedregosa2011scikit} was utilized to create the split. Every output from the quantum circuit was aggregated from 1024 shots. Lastly, to compare the results against the current classification modeling standard, the LightGBM gradient boosting algorithm \cite{ke2017lightgbm} is also trained with 50 train/test splits of each data set.

The box plots of the accuracy, F1 score, and AUC score are given in Figure \ref{fig:results}. Unsurprisingly, given the complexity of the algorithm, LightGBM consistently outperformed the quantum statistical models, as indicated by the high valued and small range of the quartiles. However, the IQP encoding method performed very close to LightGBM on every data set. Interestingly, the EntAngle consistently underperformed; given the expressivity of the feature method from the entangled layer one would at least expect the EntAngle to outperform the Angle feature map.   

In order to determine whether any of the results were statistically equivalent, a series of statistical tests were run for each dataset and each metric. First, a series of one-way ANOVA tests were run to determine if the means of all populations are equal. For all groups, the null hypothesis was rejected with p-values less than $1\mathrm{e}{-26}$. With the confirmation of statistically significant differences, the post hoc Tukey honestly significant difference (HSD) test was run. This test allows testing of all pairwise sets of means. As suggested by the graphs in Figure \ref{fig:results}, IQP and LightGBM consistently performed statistically equivalently, except for WDBC accuracy and WDBC F1 where LightGBM slightly outperformed IQP. P-values for the equivalent cases ranged from 0.605 to 1. Additionally, Angle and EntAngle consistenly performed statistically equivallently, with p-values typically ranging from 0.811 to 1. The only exception to this range is for the SIRTUIN-6 dataset's AUC scores, where the equivalence is instead confirmed with a p-value of 0.171. Despite being an alternate formulation of the IQP encoding, AltIQP only occasionally matched the performance of IQP and LightGBM, instead typically resulting in a p-value of 0.

To further illuminate the results and derive deeper insights, we calculate the expressibility of the the feature maps. Expressibility, defined by Sim, Johnson, and Aspuru-Guzik \cite{sim2019expressibility}, is a measure that compares the distribution of the fidelity of the states created from the feature map with respect to the Haar measure (probability measure); comparison is typically taken either from applying KL divergence or Jensen–Shannon divergence, which adjusts KL for symmetry. It is fairly common for the expressibility to be calculated after independent random sampling of the parameters and the fidelity calculated, and a linear spanning of the interval $[0,1]$ with respect to the number of random samples. The independence is essential for the probability distribution function of the Haar measure, which, for $f$ a fidelity of two states and $N$ is the dimension of the Hilbert space, is of the form 
\begin{equation}\label{eq:unif-haar}
    P(f) = (N-1)(1-f)^{N-2}.
\end{equation}
For applications $N=2^n$ where $n$ is the number of qubits.

Given the dependence on the data sets, to calculate the expressibility, the fidelity between all different data points is calculated and applied to the Haar measure \cite{zyczkowski2005average,sim2019expressibility}. Of course, one then losses independence between parameters and Equation \ref{eq:unif-haar} does not hold. {\.Z}yczkowski and Sommers \cite{zyczkowski2005average} give a general probability distribution function for this case,
\begin{equation}\label{eq:depend-haar}
    P_K(f) = \frac{ \Gamma(KN) }{ \Gamma(KN) \Gamma\big( K(N-1)\big) } f^{K-1}(1-f)^{k(N-1)-1}
\end{equation}
with $\Gamma$ the gamma function, and $K\cdot N$ is the dimension of the Hilbert space. 

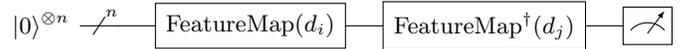
\begin{figure}[htbp]
\centering
    \begin{quantikz}[thin lines] 
       \lstick{$\ket{0}^{\otimes n}$} & \qwbundle[]{n} &  \gate{ \text{FeatureMap}(d_i)} & \gate{ \text{FeatureMap}^{\dagger}(d_j) } &\meter{}
    \end{quantikz}
\caption{\label{fig:fidelity} General circuit of the fidelity test for an arbitrary feature map for data points $d_i$ and $d_j$.}
\end{figure}

For calculations, the fidelity test \ref{fig:fidelity} is applied to get the fidelity of the states, with 1,000 shots, and the histogram function in NumPy \cite{harris2020array} is used to approximate the distribution of values; this distribution is denoted as \textbf{PQC}. The gamma function from SciPy \cite{2020SciPy-NMeth} is utilized to calculate the Haar measure where the fidelity values span the interval $(0,1)$ with respect to the bins of the histogram. The number of bins for the histogram was set to 100 and, for $n$ the number of qubits, $K=2^{n/2}$ if $n$ is even, otherwise $K^{\lfloor n/2 \rfloor}$. 

\begin{figure}[htbp]
    \begin{adjustbox}{valign=t}
    \subfigure[\label{subfig:expr-amp} Expressibility of the Amplitude feature map.
    ]{ \includegraphics[scale=.038]{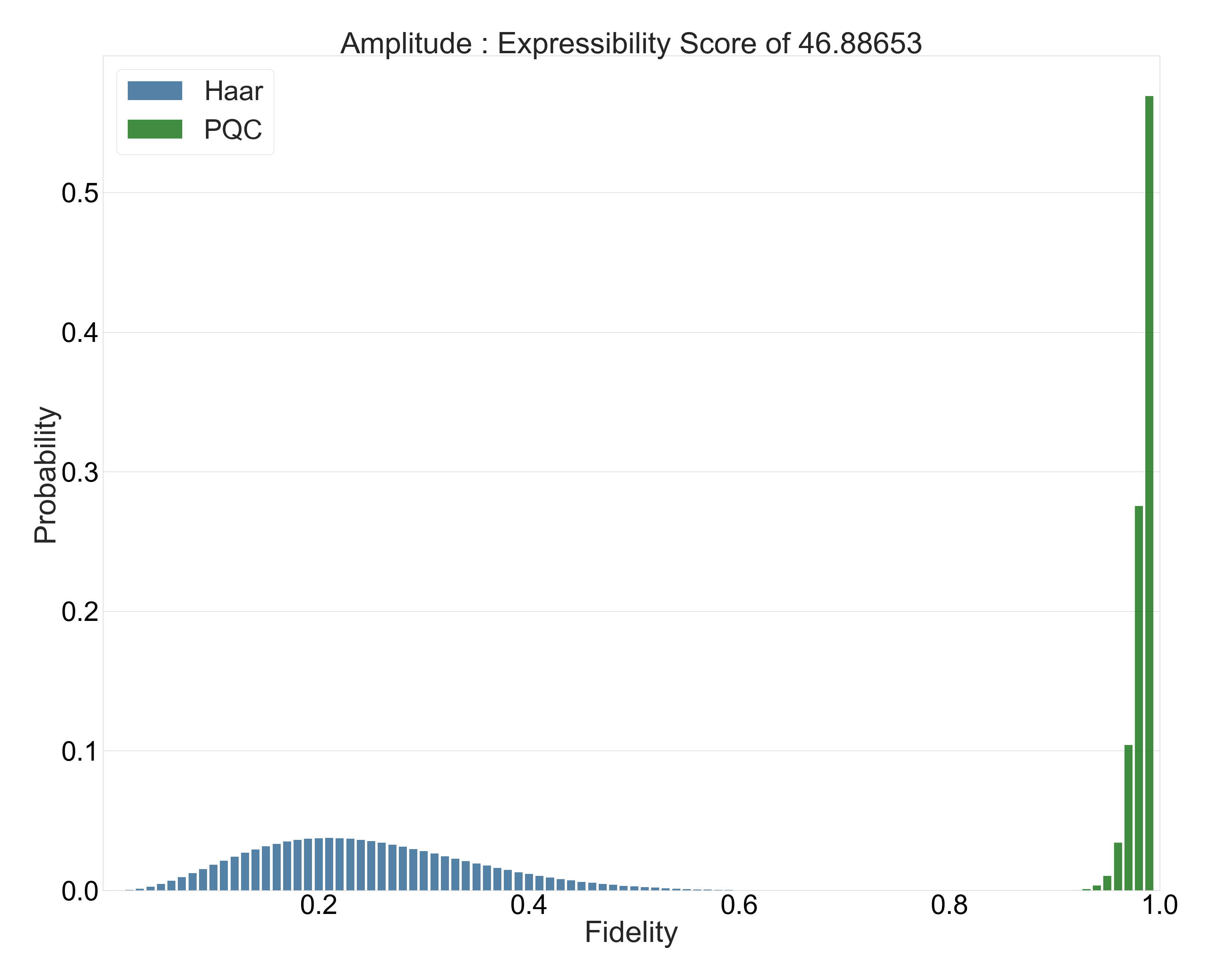}
    }
    \end{adjustbox}
    \begin{adjustbox}{valign=t} 
    \subfigure[\label{subfig:expr-angle }Expressibility of the Angle feature map.]{\includegraphics[scale=.038]{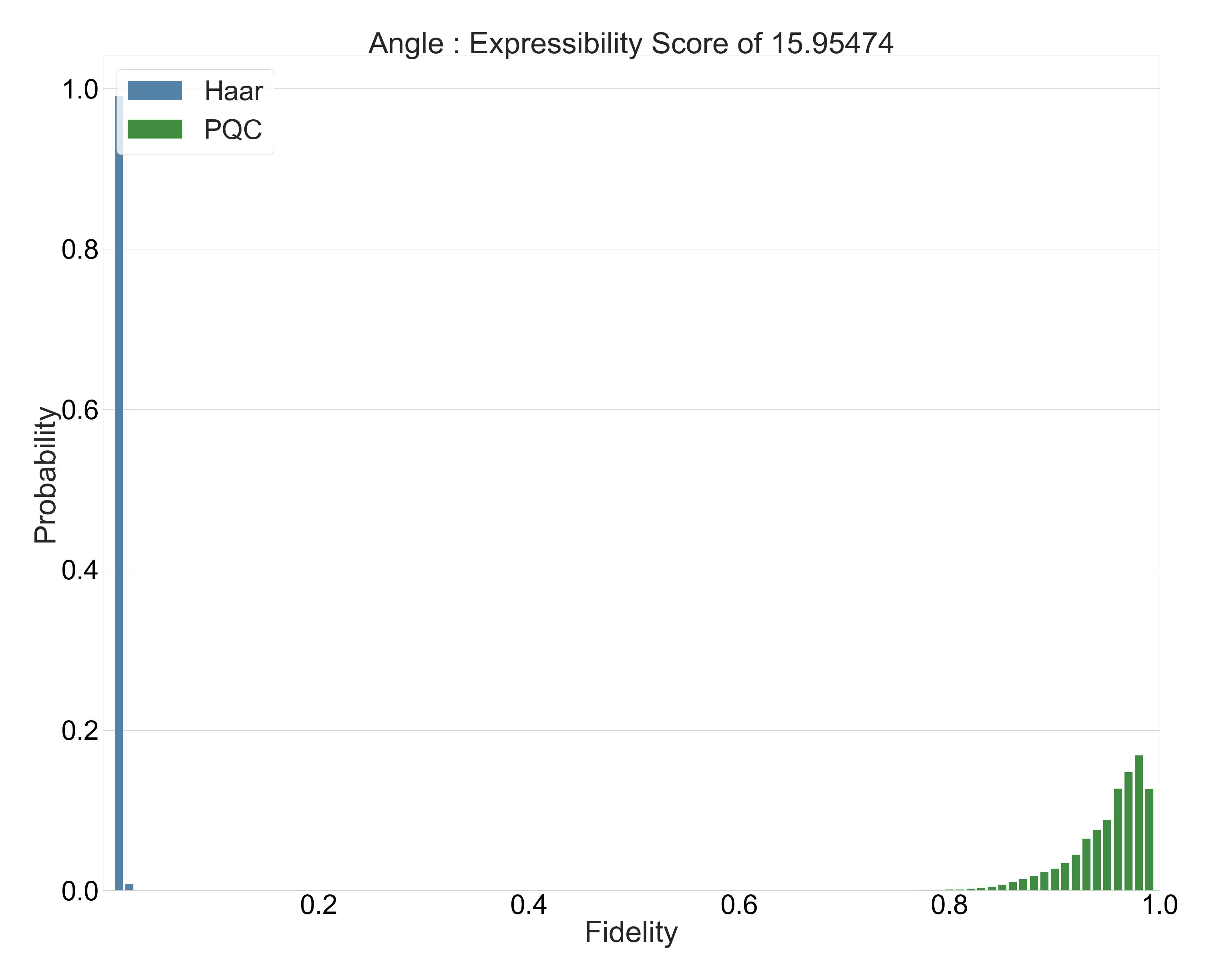} 
    }
    \end{adjustbox}
    \begin{adjustbox}{valign=t} 
    \subfigure[\label{subfig:expr-entangle}Expressibility of the EntAngle feature map.]{ \includegraphics[scale=.038]{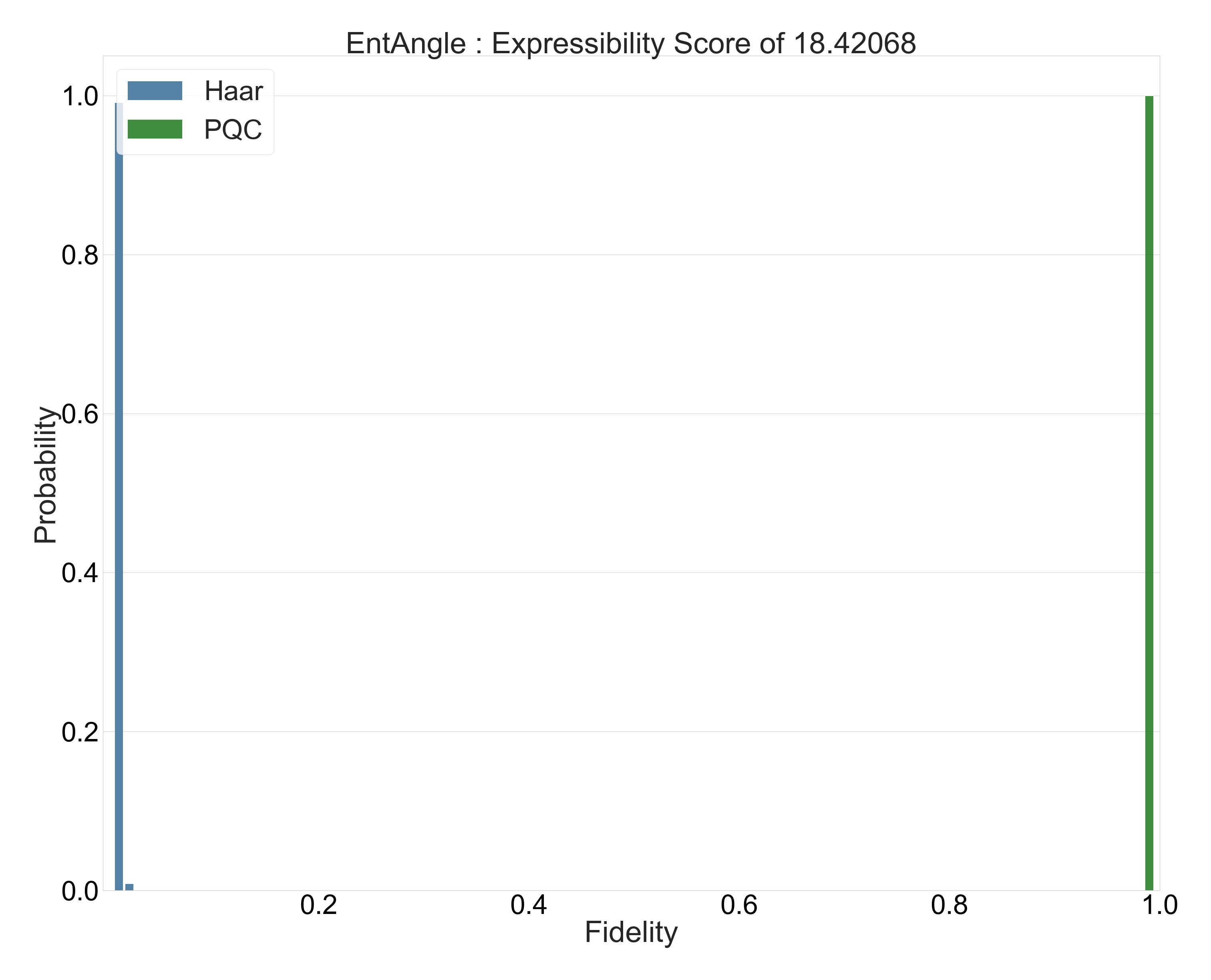}
    }
    \end{adjustbox}
    \begin{adjustbox}{valign=t} 
    \subfigure[\label{subfig:expr-iqp}Expressibility of the IQP feature map.]{ \includegraphics[scale=.038]{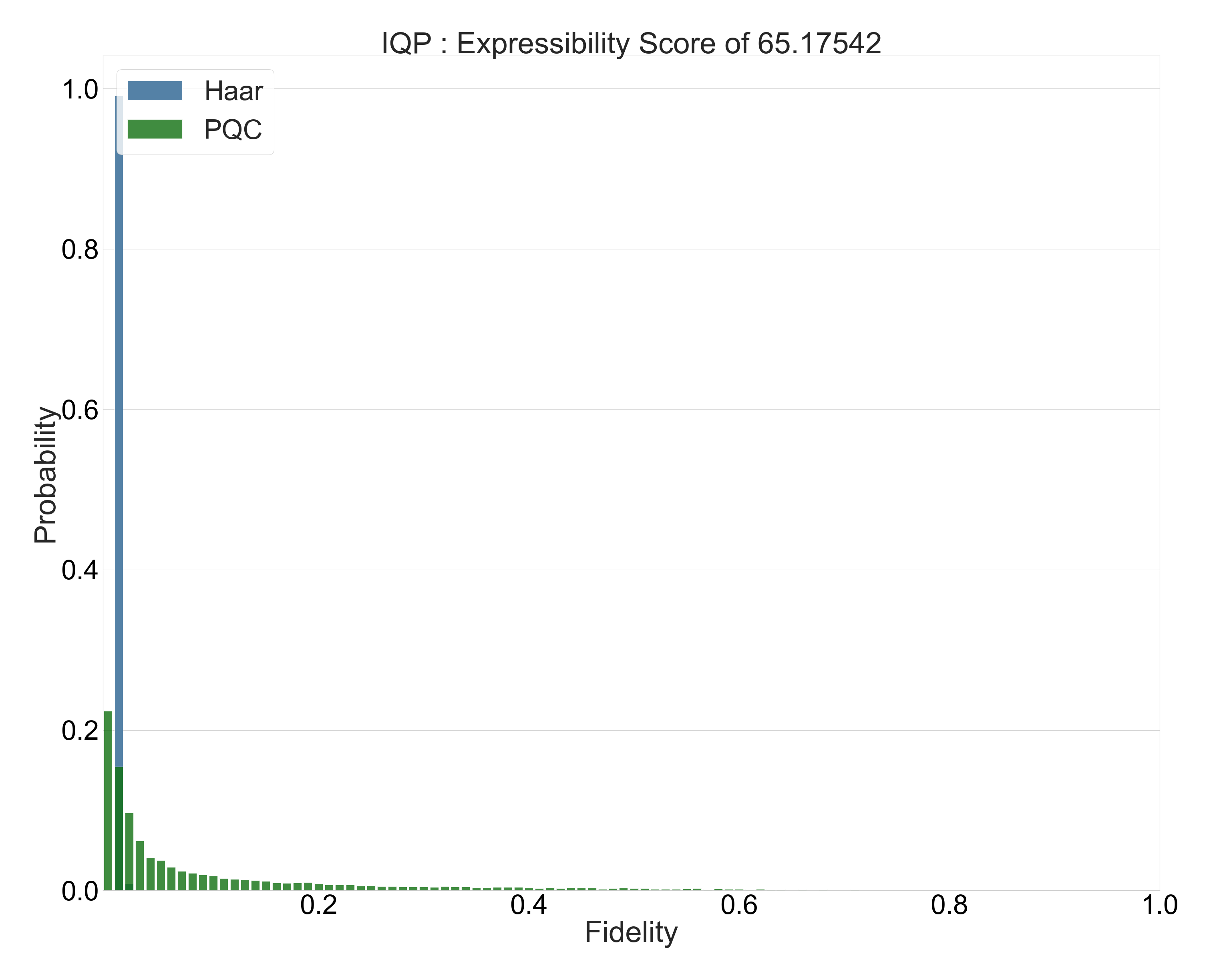}
    }
    \end{adjustbox}
    \begin{adjustbox}{valign=t} 
    \subfigure[\label{subfig:expr-altiqp}Expressibility of the AltIQP feature map.]{ \includegraphics[scale=.038]{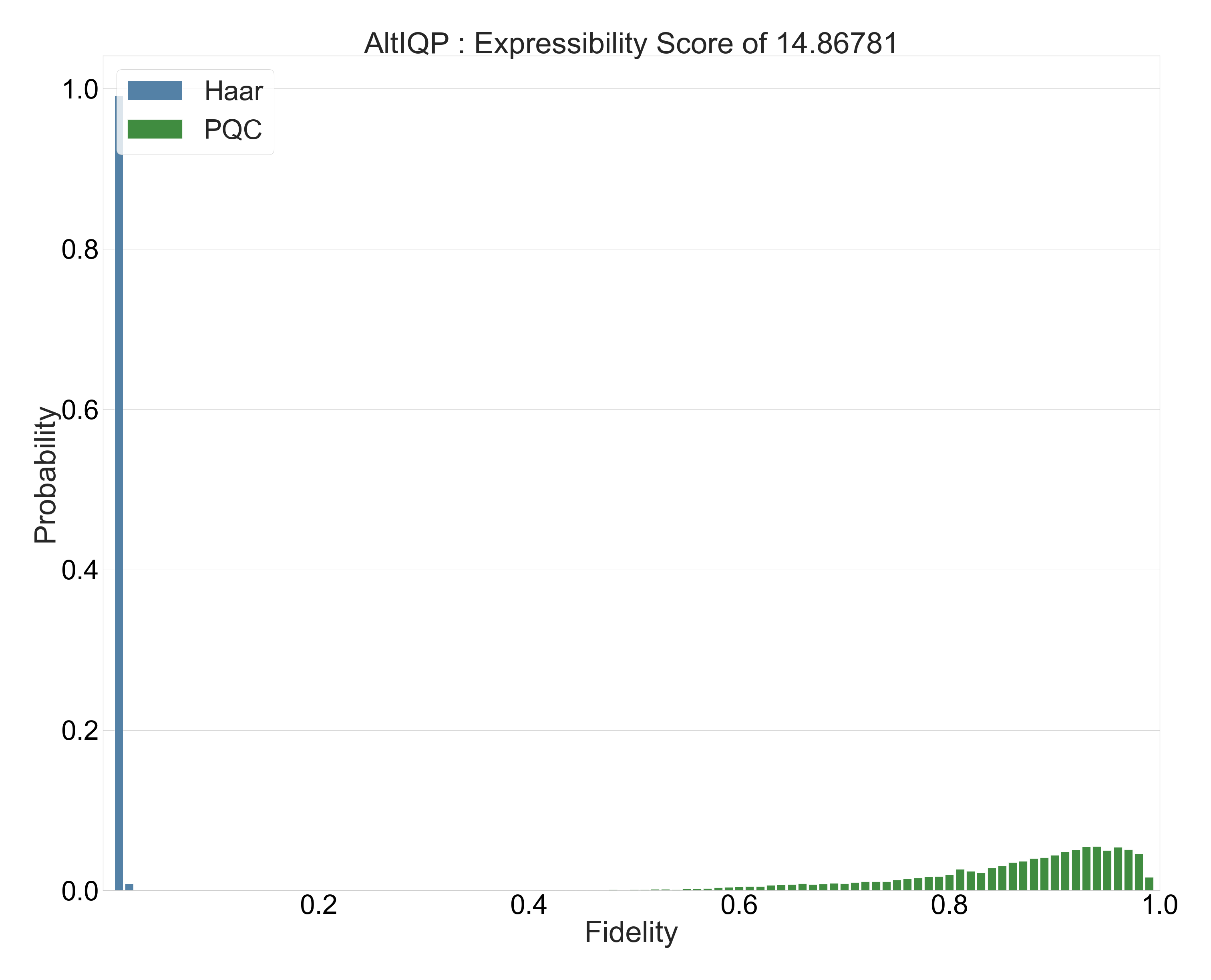}
    }
    \end{adjustbox}
\caption{\label{fig:express}The PQC and Haar distributions and expressibility score for all feature maps on the Sonar data set.}
\end{figure}

To illuminate the distributions, Figure \ref{fig:express} focuses on the Sonar data set. Observe that the distribution of the fidelities generated by the EntAngle feature map is concentrated around $1$, displaying that the different states are hard to distinguish, which would lead to a poor performing model. Given the lack of concentration of the Angle's feature map PQC distribution, this would imply the entangled layer is the reason for the ambiguity of states.

\begin{table}[hbt]
\begin{adjustbox}{width=\columnwidth}
\begin{tabular}{@{\extracolsep\fill}cccccc}
 \hline \multicolumn{6}{@{}c@{}}{\textbf{Expressibility Scores}} \\ \hline
  & Amplitude & Angle & EntAngle & IQP & AltIQP\\ \hline 
 Sonar & 46.88 & 15.95 & 18.42 & 65.17 & 14.87 \\ \hline
 Ionosphere & 27.48 & 16.09 & 18.42 & 22.12 & 56.64 \\ \hline
 WDBC & 52.59 & 16.94 & 18.42 & 93.08 & 15.87 \\ \hline
 Sirtuin6 & 23.55 & 344.82 & 605.14 & 38.83 & 170.80 \\ \hline 
 $[0,2\pi]^6$ & 2.503 & 10.401 & 667.28 &  0.579 & 5.503 \\ 
 \hline 
 $[0,2\pi]^{10}$ & 5.501 & $\infty$ & 18.421 & 0.3032 & $\infty$ \\
 \hline
\end{tabular} 
\end{adjustbox}
 \caption{ }\label{tab:compare}
\end{table}

\begin{figure}[htbp]
\includegraphics[scale=.08]{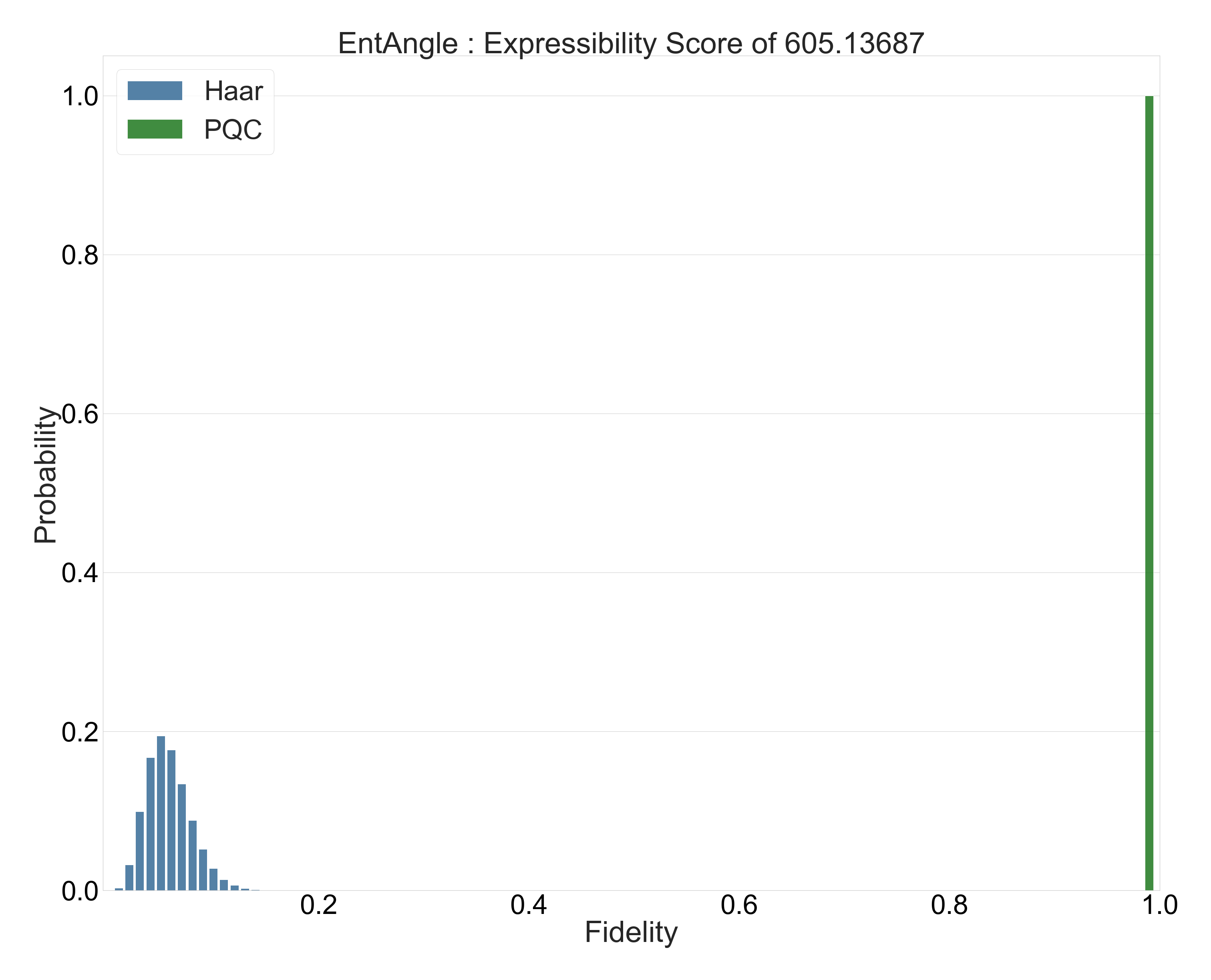}
\caption{\label{fig:6expr}The PQC and Haar distributions and expressibility score Sirtuin6 with the EntAngle feature map.}
\end{figure}

Interestingly, the smaller score for EntAngle, for all data sets except Sirtuin6, comes from the concentrations of the distributions, expanding the number of bins where both distributions have no mass. The large score of Sirtuin6, shown in Figure \ref{fig:6expr}, comes from the Haar distribution having mass in more bins, stemming from the number of features leading to a smaller number of qubits. The mass of fidelity is still around 1. 

To display the difference of the data with respect to expressibility over the entire space, the expressibility of the feature maps with random data points of six dimensions and 10 dimensions is calculated and set in Table \ref{tab:compare}. Since the random sampling is independent, the Haar measure from Equation \ref{eq:unif-haar} is applied. For the sampling, two random data sets with 150 are sampled from the $[0,1]^6$ and $[0,1]^{10}$, respectively, then every array is multiplied by $2\pi$, using Numpy package \cite{harris2020array}. 

The computation was conducted with Arizona State University's supercomputer \cite{HPC:ASU23}

\section{\label{sec:discussion}Discussion}
This work showcased two primary results that are of particular interest. It was unexpected for the Angle and EntAngle approaches to perform equivalently. Entanglement is one of the strengths of quantum computing, and enables enhancement of quantum performance \cite{baumer2021demonstrating}. However, the application of entangling gates had a negligible impact on the classification power of the model. This consistency is attributed in large part to the architecture of the QSVC. Although the model employs quantum encodings of data, that component only serves as an analog to a kernel. The distance is extracted through a measurement, collapsing the entanglement, and feeding the data to the classical SVC component. As such, we suggest that these results are due to the entanglement occurring immediately before a measurement and not varying between data points in a way that would enhance classifiability.

\begin{figure}[htbp]
    \begin{adjustbox}{valign=t}
    \subfigure[\label{subfig:expr-unif-6amp} Expressibility of the Amplitude feature map for six dimensions.
    ]{ \includegraphics[scale=.038]{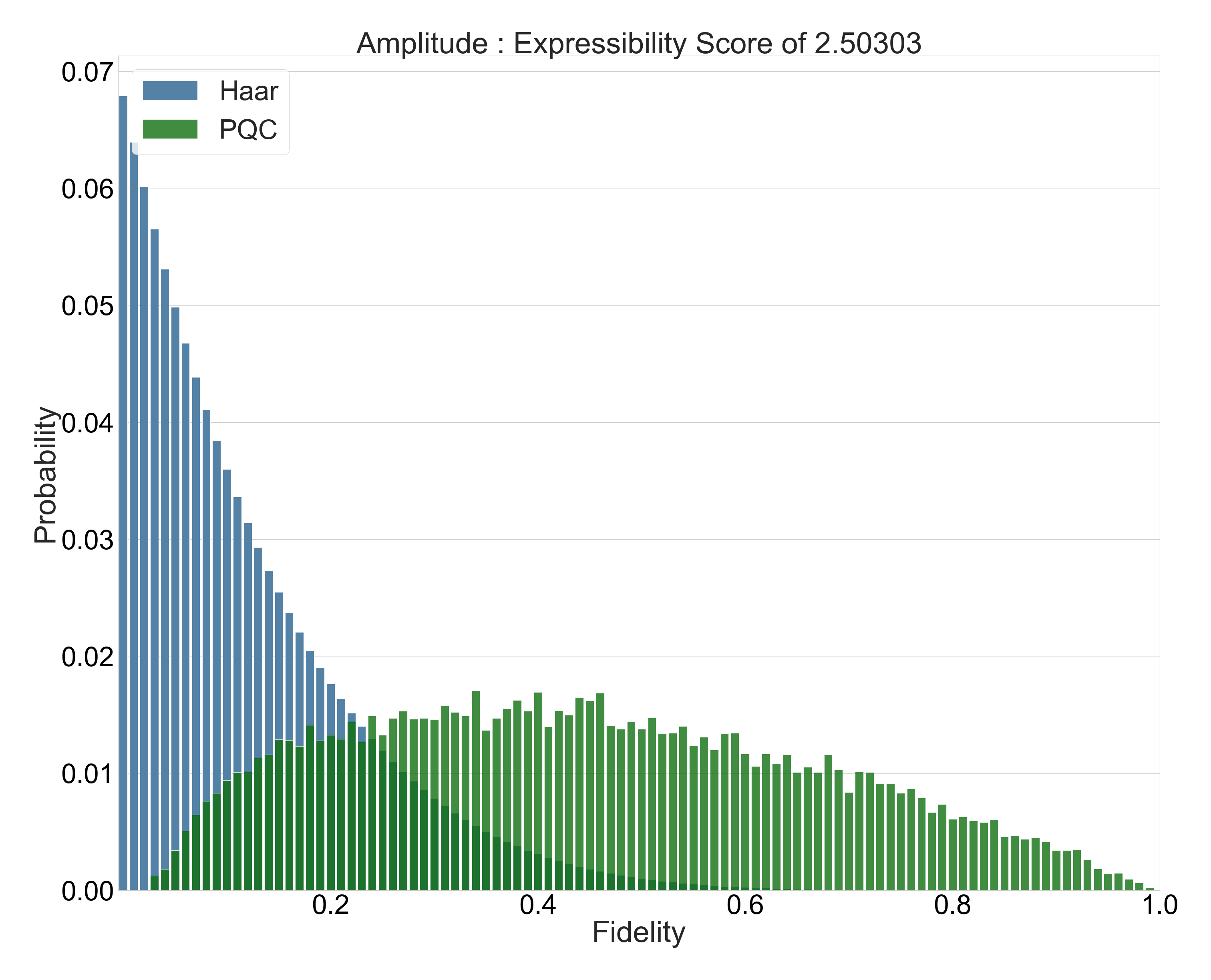}
    }
    \end{adjustbox}
    \begin{adjustbox}{valign=t} 
    \subfigure[\label{subfig:expr-unif-10amp}Expressibility of the Amplitude feature map for ten dimensions.]{\includegraphics[scale=.038]{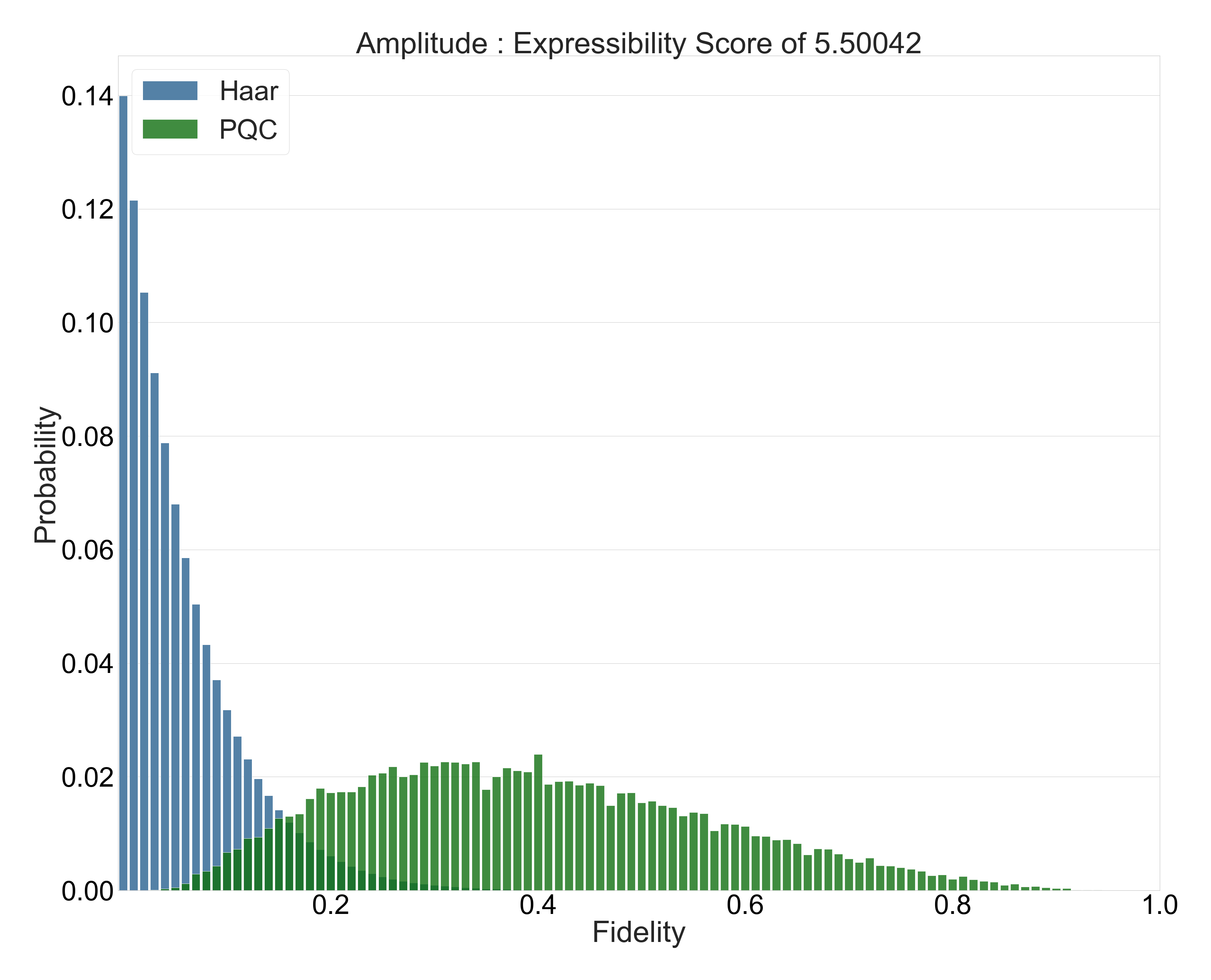} 
    }
    \end{adjustbox}
    
    \begin{adjustbox}{valign=t} 
    \subfigure[\label{subfig:expr-unif-6angle}Expressibility of the Angle feature map for six dimensions.]{ \includegraphics[scale=.038]{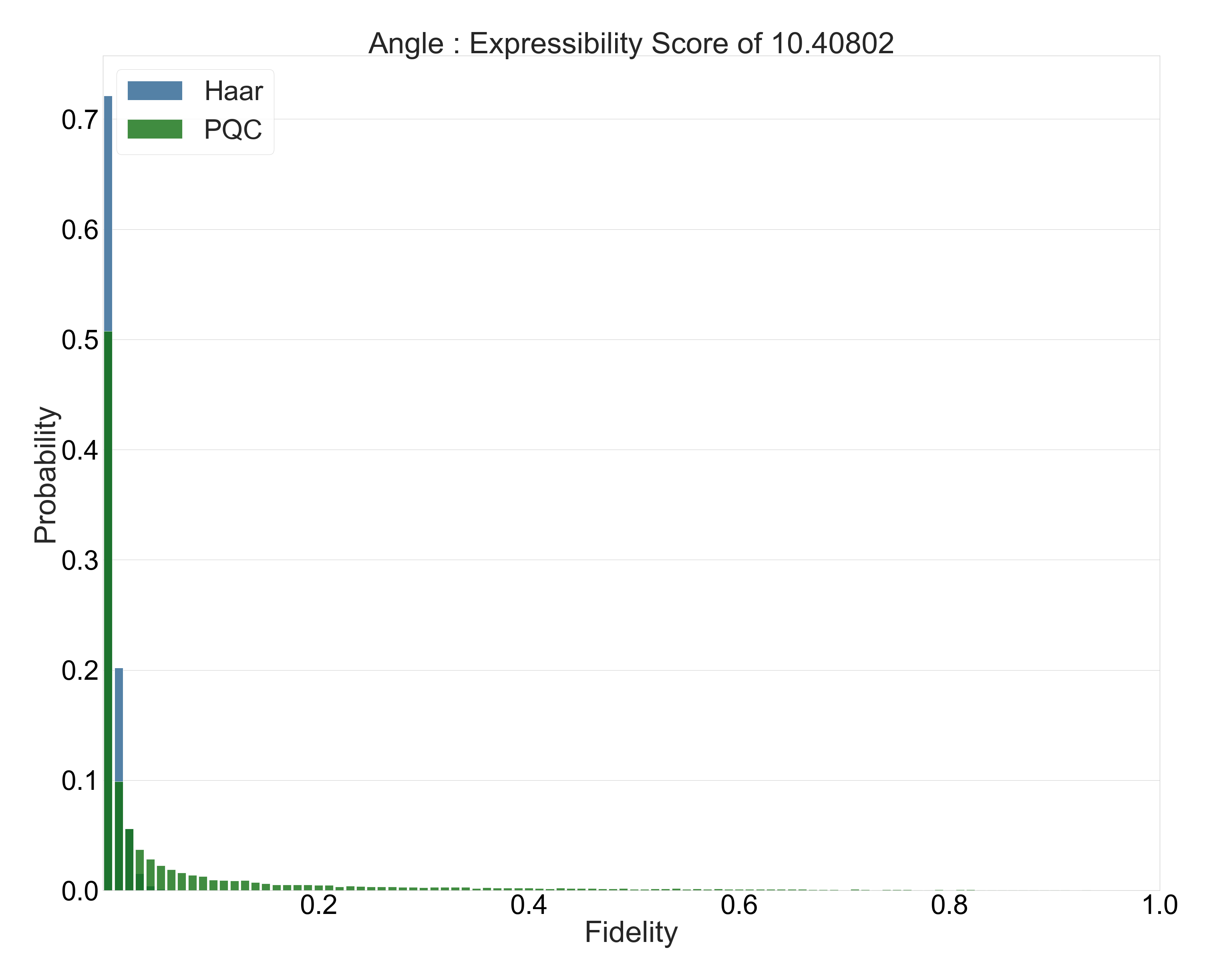}
    }
    \end{adjustbox}
    \begin{adjustbox}{valign=t} 
    \subfigure[\label{subfig:expr-unif-10angle}Expressibility of the Angle feature map for ten dimensions.]{ \includegraphics[scale=.038]{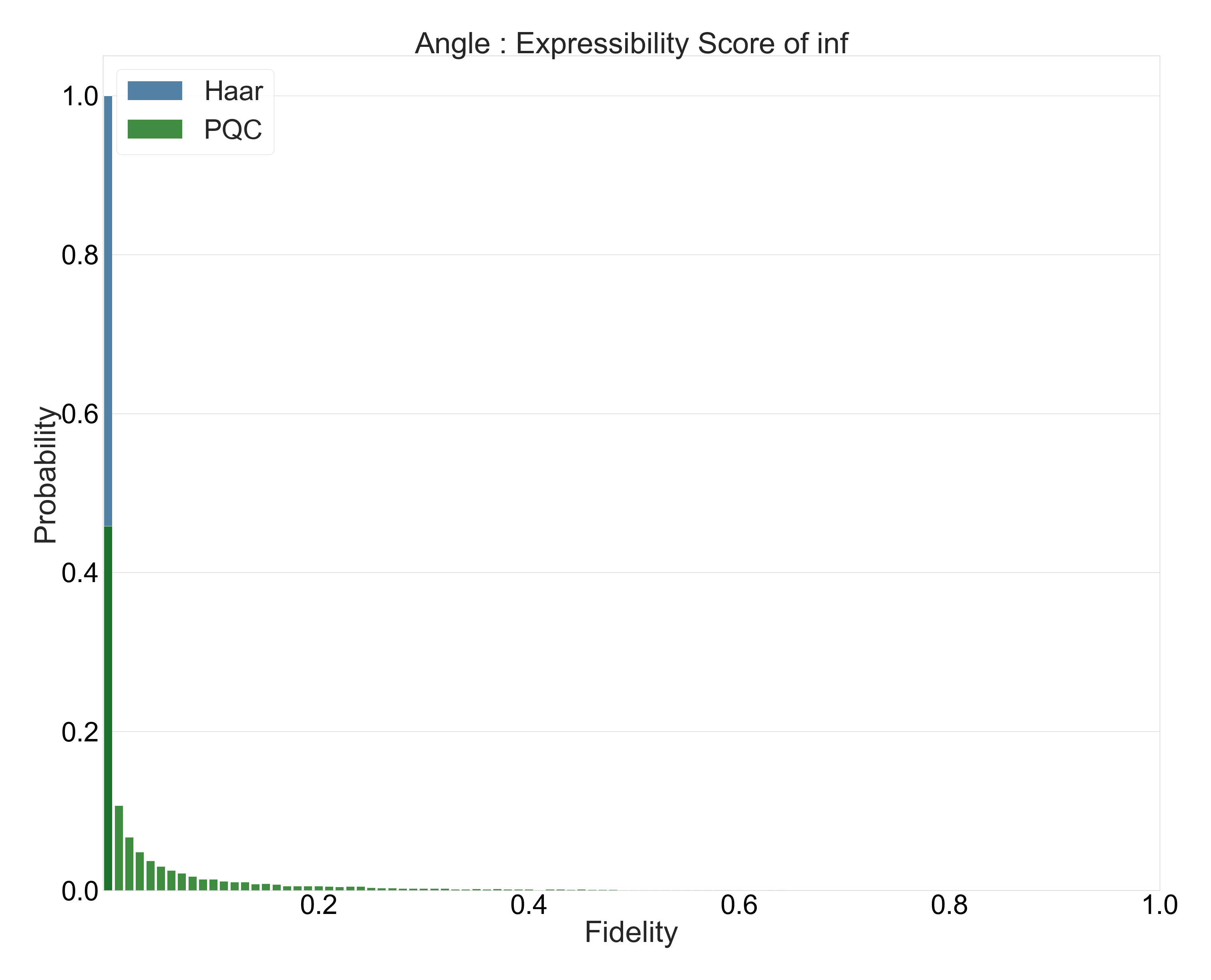}
    }
    \end{adjustbox}
    
    \begin{adjustbox}{valign=t} 
    \subfigure[\label{subfig:expr-unif-6entangle}Expressibility of the EntAngle feature map for six dimensions.]{ \includegraphics[scale=.038]{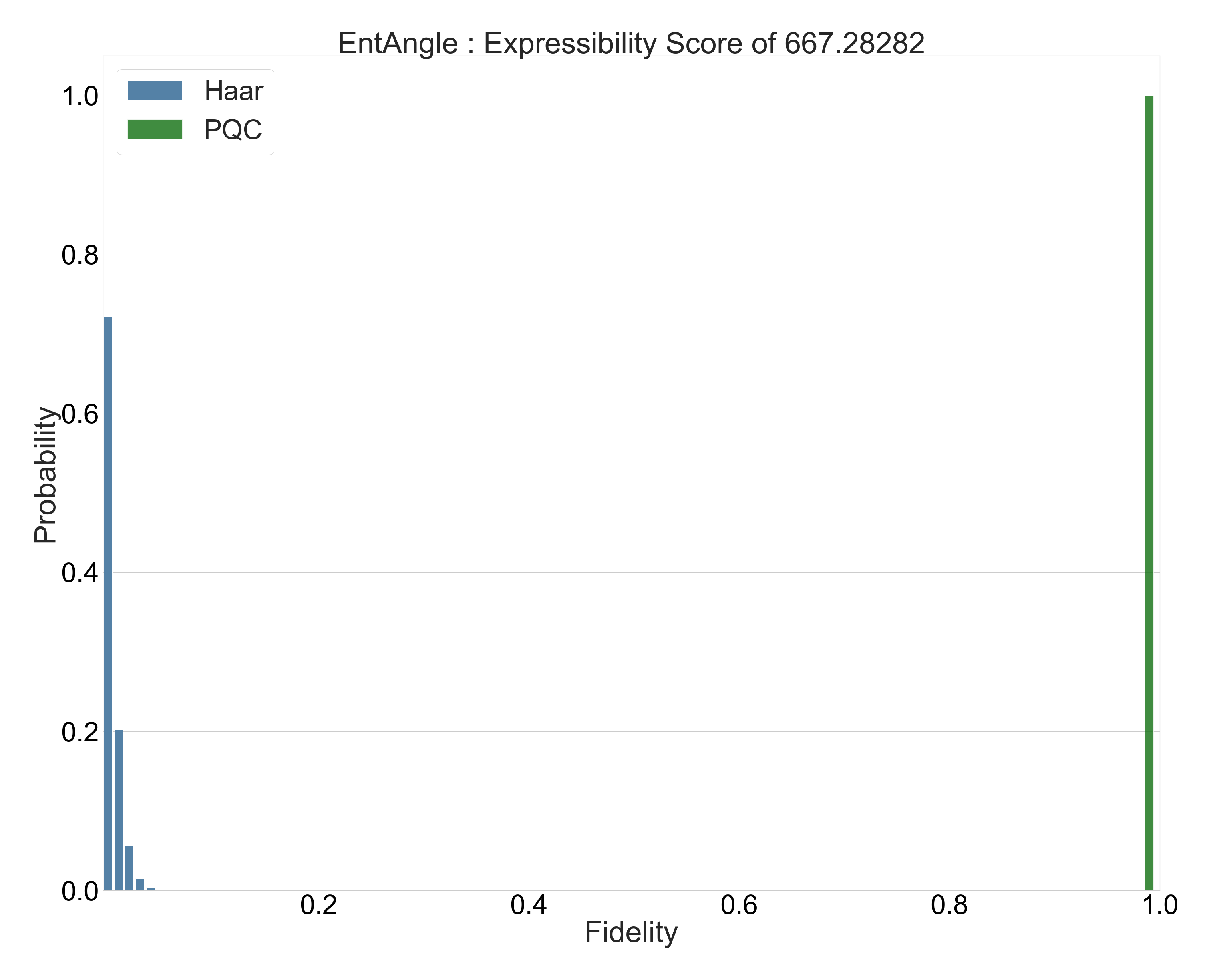}
    }
    \end{adjustbox}
    \begin{adjustbox}{valign=t} 
    \subfigure[\label{subfig:expr-unif-10entangle}Expressibility of the EntAngle feature map for 10 dimensions.]{ \includegraphics[scale=.038]{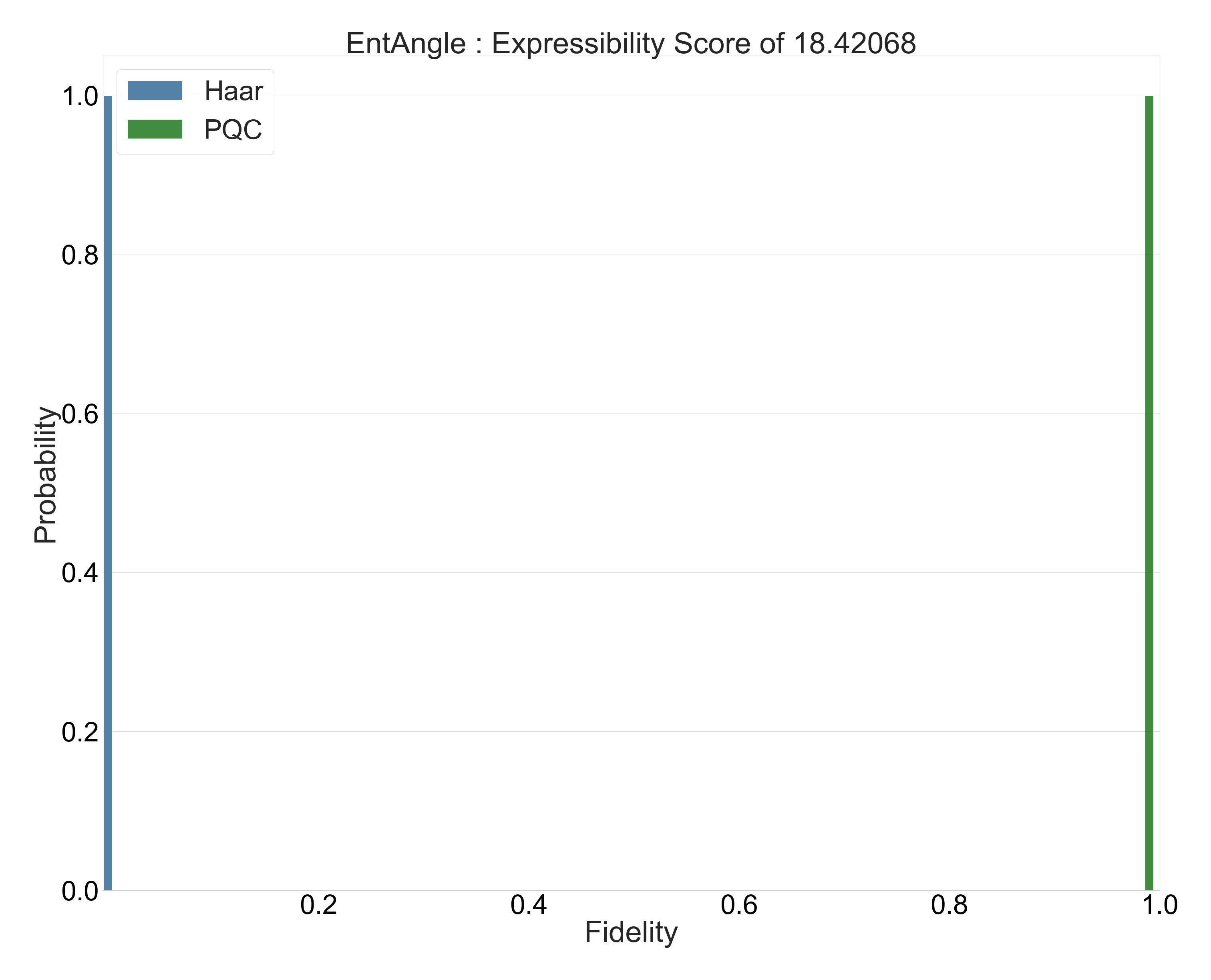}
    }
    \end{adjustbox}

    \begin{adjustbox}{valign=t} 
    \subfigure[\label{subfig:expr-unif-6iqp}Expressibility of the IQP feature map for six dimensions.]{ \includegraphics[scale=.038]{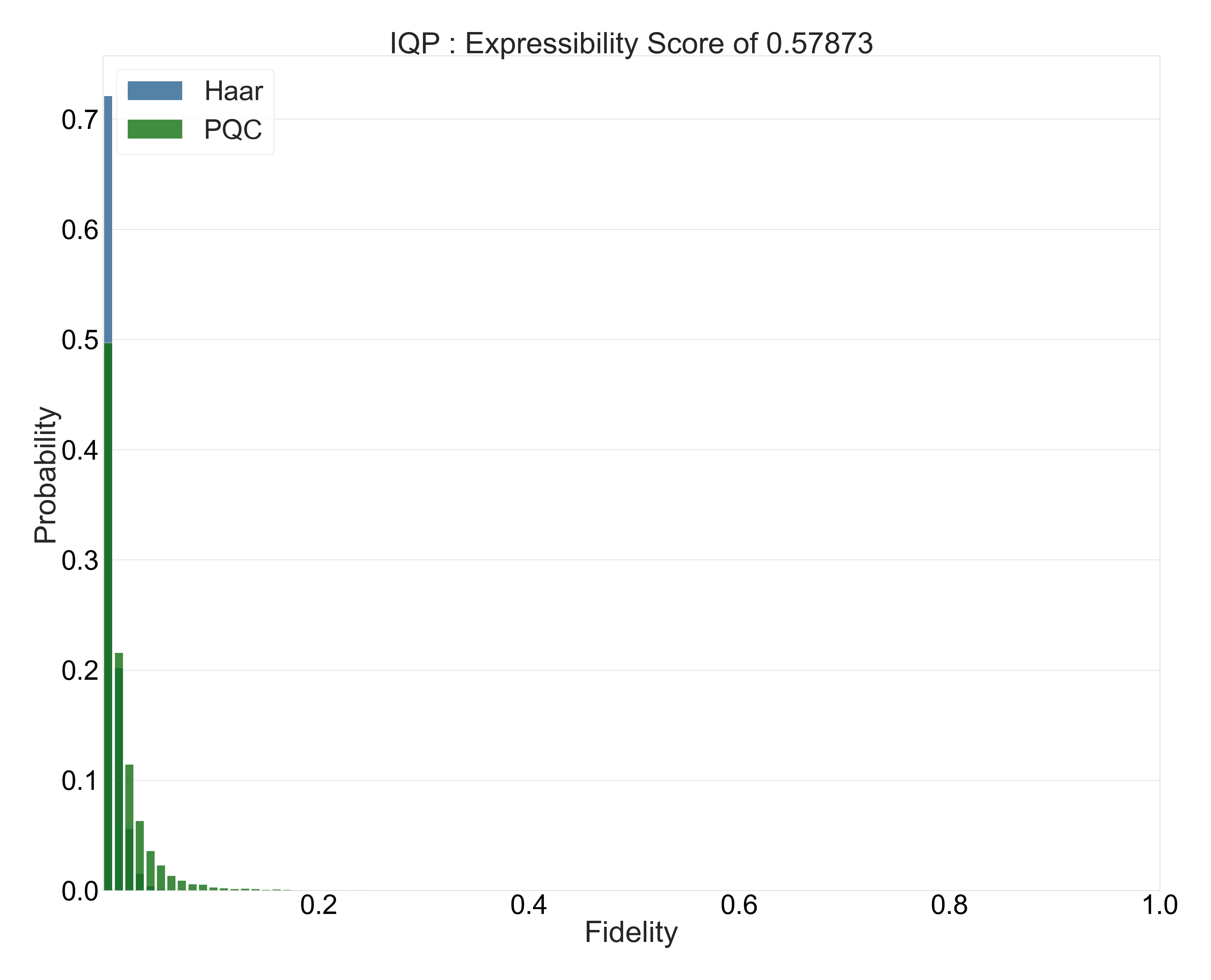}
    }
    \end{adjustbox}
    \begin{adjustbox}{valign=t} 
    \subfigure[\label{subfig:expr-unif-10iqp}Expressibility of the IQP feature map  for ten dimensions.]{ \includegraphics[scale=.038]{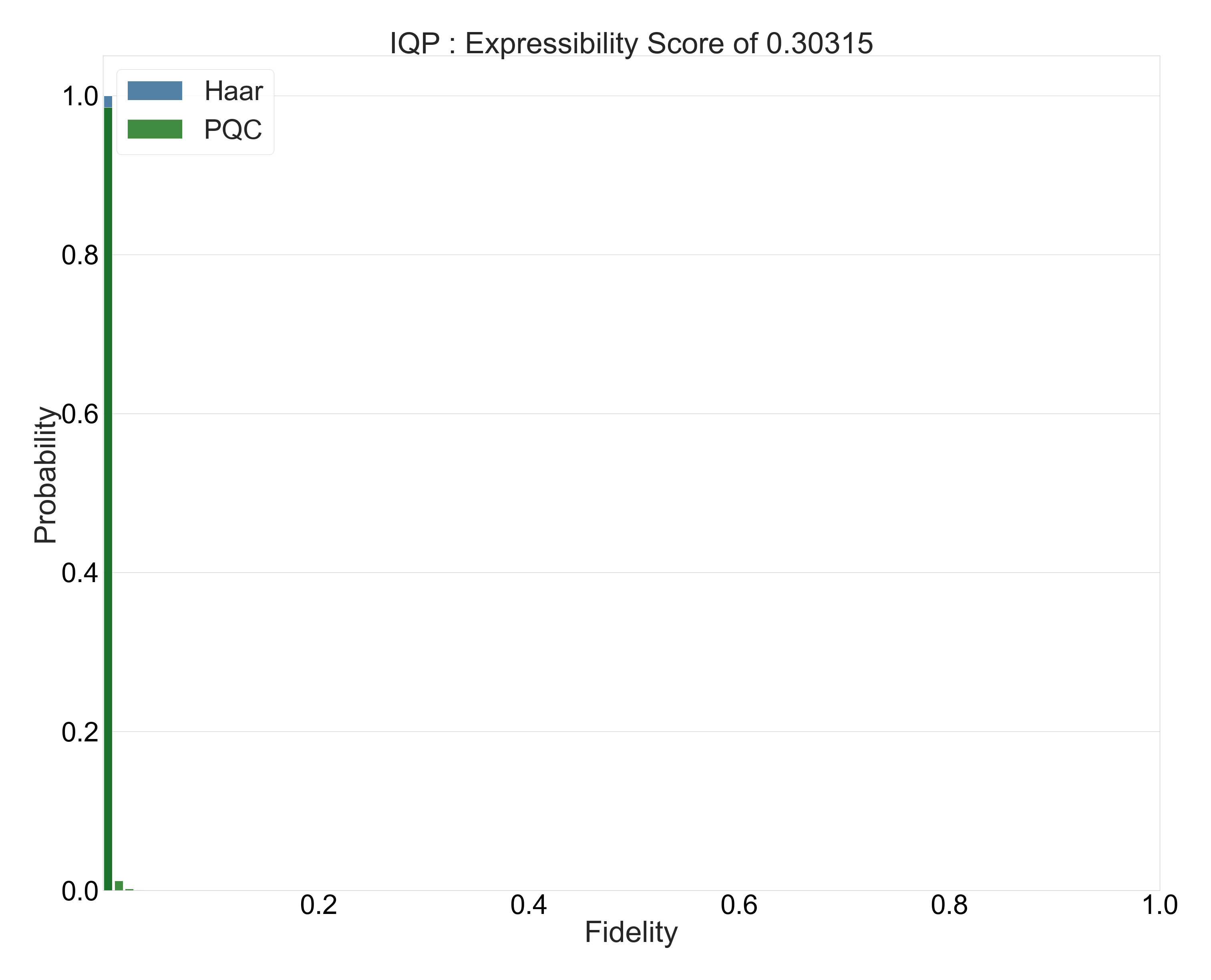}
    }
    \end{adjustbox}

    \begin{adjustbox}{valign=t} 
    \subfigure[\label{subfig:expr-unif-6altiqp}Expressibility of the AltIQP feature map for six dimensions.]{ \includegraphics[scale=.038]{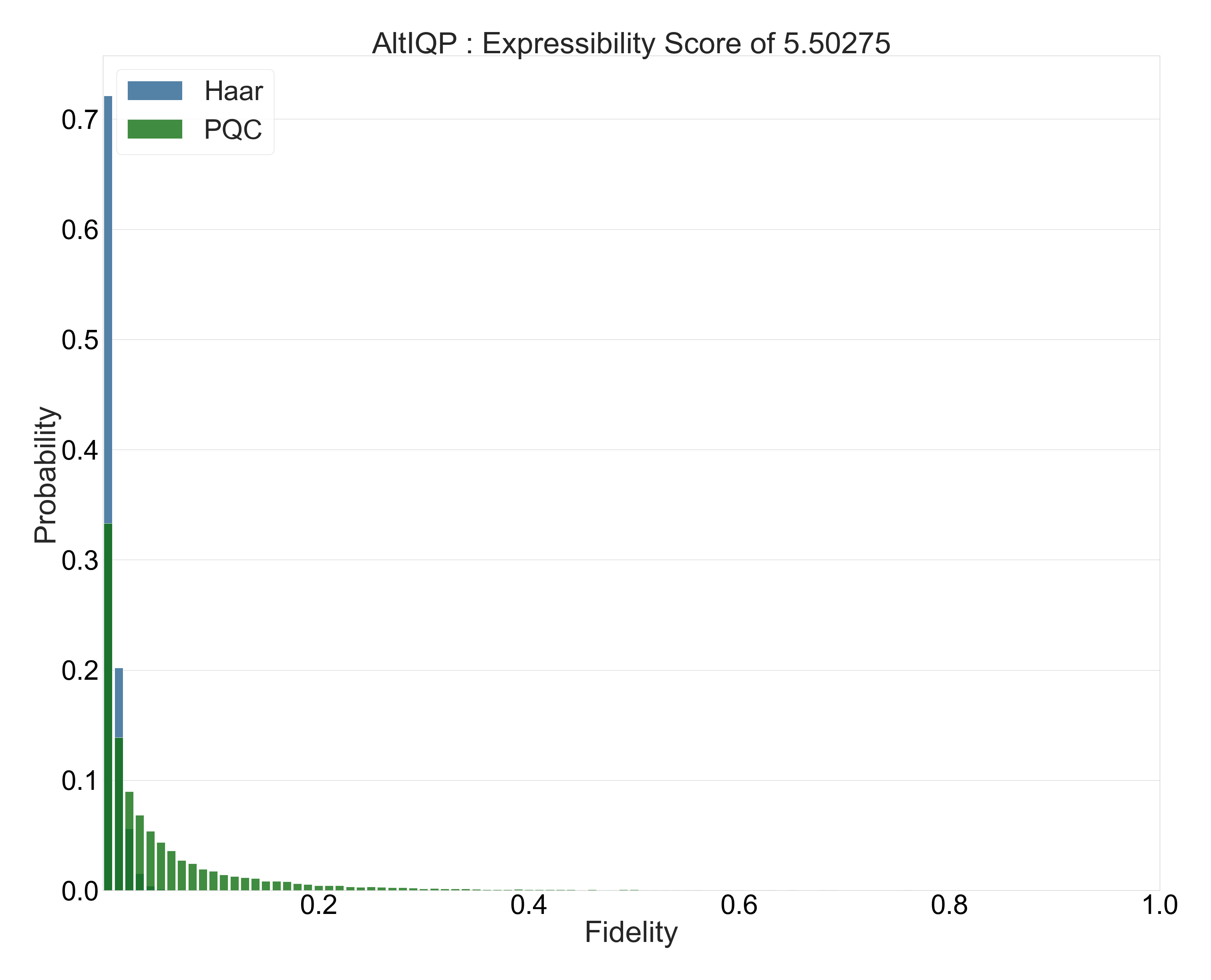}
    }
    \end{adjustbox}
    \begin{adjustbox}{valign=t} 
    \subfigure[\label{subfig:expr-unif-10altiqp}Expressibility of the AltIQP feature map for ten dimensions.]{ \includegraphics[scale=.038]{figures/ExpressUnif_Angle_10.png}
    }
    \end{adjustbox}
\caption{\label{fig:express-unif}The PQC and Haar distributions and expressibility score for all feature maps on the Sonar data set.}
\end{figure}

The other major result of our work is the statistically equivalent performance of the IQP encoding and the classical LightGBM model. One of the major advantages of quantum machine learning is its ability to generalize well from fewer training samples than classical approaches \cite{caro2022generalization}. The ability of the models to perform on par with a classical approach suggests an area of quantum advantage, namely the use case where classical performance would be sufficiently high, but the amount of data is not sufficient to effectively train a classical model. Of additional interest is the fact that the encoding that performed well belongs to the IQP class, which is believed to be difficult to simulate classically \cite{bremner2016average}, further suggesting some advantage to using the quantum approach.

Future work on this work may include further experimentation on the effect of entangling data within the QSVC architecture. Though this work suggested no effect, entanglement was not explored in depth at different points in the circuit or for different encoding strategies. Another area of future work is a deeper analysis of the efficacy of IQP class encoding strategies. In the two approaches explored in this work, IQP performed on par with LightGBM, whereas AltIQP failed to consistently match that performance. Work may additionally be done into identifying aspects of the IQP encoding architecture that are well suited to the targeted datasets.

\section{\label{sec:availability}Data Availability}
The experiment results and full statistical test results are available upon request.

\section{Disclaimer}
This preprint has not undergone peer review or any post-submission improvements or corrections. The Version of Record of this article is published in Quantum Machine Intelligence, and is available online at https://doi.org/10.1007/s42484-025-00298-w 

\medskip
About Deloitte: Deloitte refers to one or more of Deloitte Touche Tohmatsu Limited, a UK private company limited by guarantee (“DTTL”), its network of member firms, and their related entities. DTTL and each of its member firms are legally separate and independent entities. DTTL (also referred to as “Deloitte Global”) does not provide services to clients. In the United States, Deloitte refers to one or more of the US member firms of DTTL, their related entities that operate using the “Deloitte” name in the United States and their respective affiliates. Certain services may not be available to attest clients under the rules and regulations of public accounting. Please see  www.deloitte.com/about to learn more about our global network of member firms.

Deloitte provides industry-leading audit, consulting, tax and advisory services to many of the world’s most admired brands, including nearly 90\% of the Fortune 500® and more than 8,500 U.S.-based private companies. At Deloitte, we strive to live our purpose of making an impact that matters by creating trust and confidence in a more equitable society. We leverage our unique blend of business acumen, command of technology, and strategic technology alliances to advise our clients across industries as they  build their future. Deloitte is proud to be part of the largest global professional services network serving our clients in the markets that are most important to them. Bringing more than 175 years of service, our network of member firms spans more than 150 countries and territories. Learn how Deloitte’s approximately 457,000 people worldwide connect for impact at  www.deloitte.com.

This publication contains general information only and Deloitte is not, by means of this [publication or presentation], rendering accounting, business, financial, investment, legal, tax, or other professional advice or services. This [publication or presentation] is not a substitute for such professional advice or services, nor should it be used as a basis for any decision or action that may affect your business. Before making any decision or taking any action that may affect your business, you should consult a qualified professional advisor.
Deloitte shall not be responsible for any loss sustained by any person who relies on this publication. Copyright © 2024 Deloitte Development LLC. All rights reserved. 

\nocite{*}
\bibliography{refs}

\end{document}